\documentclass[fleqn]{article} % for journal submissions
\usepackage{graphicx,url}
\usepackage[utf8x]{inputenc}
\usepackage{longtable,color}
\usepackage[top=3cm, bottom=3cm, left=3cm, right=2cm]{geometry}
\usepackage{verbatim}
\usepackage{color}
\usepackage{xspace}
\usepackage{rotating}
\usepackage{times}
\usepackage{latexsym}
\usepackage{amstext}
\usepackage{enumerate}
\usepackage{hyperref}
\usepackage{booktabs} % For formal tables
\usepackage{amssymb}

\usepackage[utf8x]{inputenc}
\usepackage{color}
\definecolor{codegreen}{rgb}{0,0.4,0}
\definecolor{codeblue}{rgb}{0,0.5,1}
\definecolor{codegray}{rgb}{0.5,0.5,0.5}
\definecolor{codepurple}{rgb}{0.58,0,0.82}
\definecolor{backcolour}{rgb}{0.95,0.95,0.92}

\usepackage[noend, vlined, linesnumbered, ruled, english]{algorithm2e}

\usepackage[scaled=0.9]{DejaVuSansMono}
\usepackage[T1]{fontenc}

\usepackage{listings}
\lstdefinestyle{mystyle}{
    commentstyle=\color{codegreen},
    keywordstyle=\color{magenta},
    numberstyle=\tiny\color{codegray},
    stringstyle=\color{codepurple},
    basicstyle=\small\ttfamily,
    breakatwhitespace=false,         
    breaklines=true,                 
    captionpos=b,                    
    keepspaces=true,                 
    showspaces=false,                
    showstringspaces=false,
    showtabs=false,                 
    tabsize=2
}
\lstdefinelanguage{SwiftK}
{
	alsoletter={<, :, >, /},
	morekeywords={app, stdout, foreach, in, filename},
    basicstyle=\footnotesize\ttfamily,
    keywordstyle=\color{codeblue},
    numberstyle=\tiny\color{codegray},
    stringstyle=\color{codepurple},
	morestring=[b]"
}

\title{A Survey of Biodiversity Informatics: Concepts, Practices, and Challenges}

\author{Luiz M. R. Gadelha Jr.$^1$\footnote{E-mail address: {\tt lgadelha@lncc.br}} \and Pedro C. de Siracusa$^1$ \and Artur Ziviani$^1$ \and Eduardo Couto Dalcin$^2$ \and Helen Michelle Affe$^2$ \and Marinez Ferreira de Siqueira$^2$ \and Luís Alexandre Estevão da Silva$^2$ \and Douglas A. Augusto$^3$ \and Eduardo Krempser$^3$ \and Marcia Chame$^3$ \and Raquel Lopes Costa$^4$ \and Pedro Milet Meirelles$^5$ and Fabiano Thompson$^6$
}
\date{\small
$\mbox{}^1$National Laboratory for Scientific Computing, Petrópolis, Brazil\\
$\mbox{}^2$Friedrich-Schiller-University Jena, Jena, Germany\\
$\mbox{}^2$Rio de Janeiro Botanical Garden, Rio de Janeiro, Brazil\\
$\mbox{}^3$Oswaldo Cruz Foundation, Rio de Janeiro, Brazil\\
$\mbox{}^4$National Institute of Cancer, Rio de Janeiro, Brazil\\
$\mbox{}^5$Federal University of Bahia, Salvador, Brazil\\
$\mbox{}^6$Federal University of Rio de Janeiro, Rio de Janeiro, Brazil\\
}

\begin{document}

\flushbottom
\maketitle

\begin{abstract}
The unprecedented size of the human population, along with its associated economic activities, have an ever increasing impact on global environments. Across the world, countries are concerned about the growing resource consumption and the capacity of ecosystems to provide them. 
To effectively conserve biodiversity, it is essential to make indicators and knowledge openly available to decision-makers in ways that they can effectively use them. The development and deployment of mechanisms to produce these indicators depend on having access to trustworthy data from field surveys and automated sensors, biological collections, molecular data, and historic academic literature. The transformation of this raw data into synthesized information that is fit for use requires going through many refinement steps. The methodologies and techniques used to manage and analyze this data comprise an area often called {\em biodiversity informatics} (or {\em e-Biodiversity}). 
Biodiversity data follows a life cycle consisting of planning, collection, certification, description, preservation, discovery, integration, and analysis. 
Researchers, whether producers or consumers of biodiversity data, will likely perform activities related to at least one of these steps. This article explores each stage of the life cycle of biodiversity data, discussing its methodologies, tools, and challenges. 
\end{abstract}

\section{Introduction}

Humanity is increasingly influencing global environments \cite{Newbold2015}. In many countries, it has raised governmental concern about the unbalance between resource consumption by these activities and the capacity of ecosystems to provide resources. This unbalance has resulted, for instance, in loss of forest cover in many places, extinction of species, and decreased availability of fresh water. Humans rely on {\em ecosystem services} in various activities. These services, such as food and water, are a result of processes that occur within these ecosystems. Various studies show that there is a strong relationship between human activities, global changes, biodiversity, ecosystem processes, and ecosystem services. Chapin et al. \cite{Chapin2000} observe that biodiversity variables, such as the number of species present, the number of individuals of each species, and which species are present, along with the types of interactions (e.g. trophic, competitive, symbiotic) that are taking place between these species, determine the {\em species traits} that affect ecosystem processes. These traits can be defined as characteristics or attributes of species that are expressed by genes or affected by the environment. They also observe that global changes, often triggered by humans, such as invasive species, increased atmospheric carbon dioxide, and land-use change can significantly alter these biodiversity variables and, consequently, the expression of species traits. This, in turn, affects ecosystem processes and their resulting services, which can have negative impacts on human development. Changes in these ecosystem services that are due to changes in biodiversity can sometimes be non-linear and stochastic, which can pose a significant risk to humans. Similar conclusions have been reached in other surveys on the relationship between biodiversity, ecosystem functioning, and ecosystem services \cite{Cardinale2012,Hooper2012}. Cardinale et al. \cite{Cardinale2012} observe that after a species become extinct, the resulting changes to ecological processes strongly depend on which traits were eliminated. Hooper et al. \cite{Hooper2012} observe that biodiversity loss is as significant to ecosystem change as the direct effects of global changes, such as elevated carbon dioxide in the atmosphere and ozone depletion.  
This, in turn, affects critical ecosystem services for the local population, such as food production, air quality, and fresh water. 

% --------------
% --- Figure ---
% --------------
\begin{figure*}[h]
\begin{center}
\includegraphics[width=12cm]{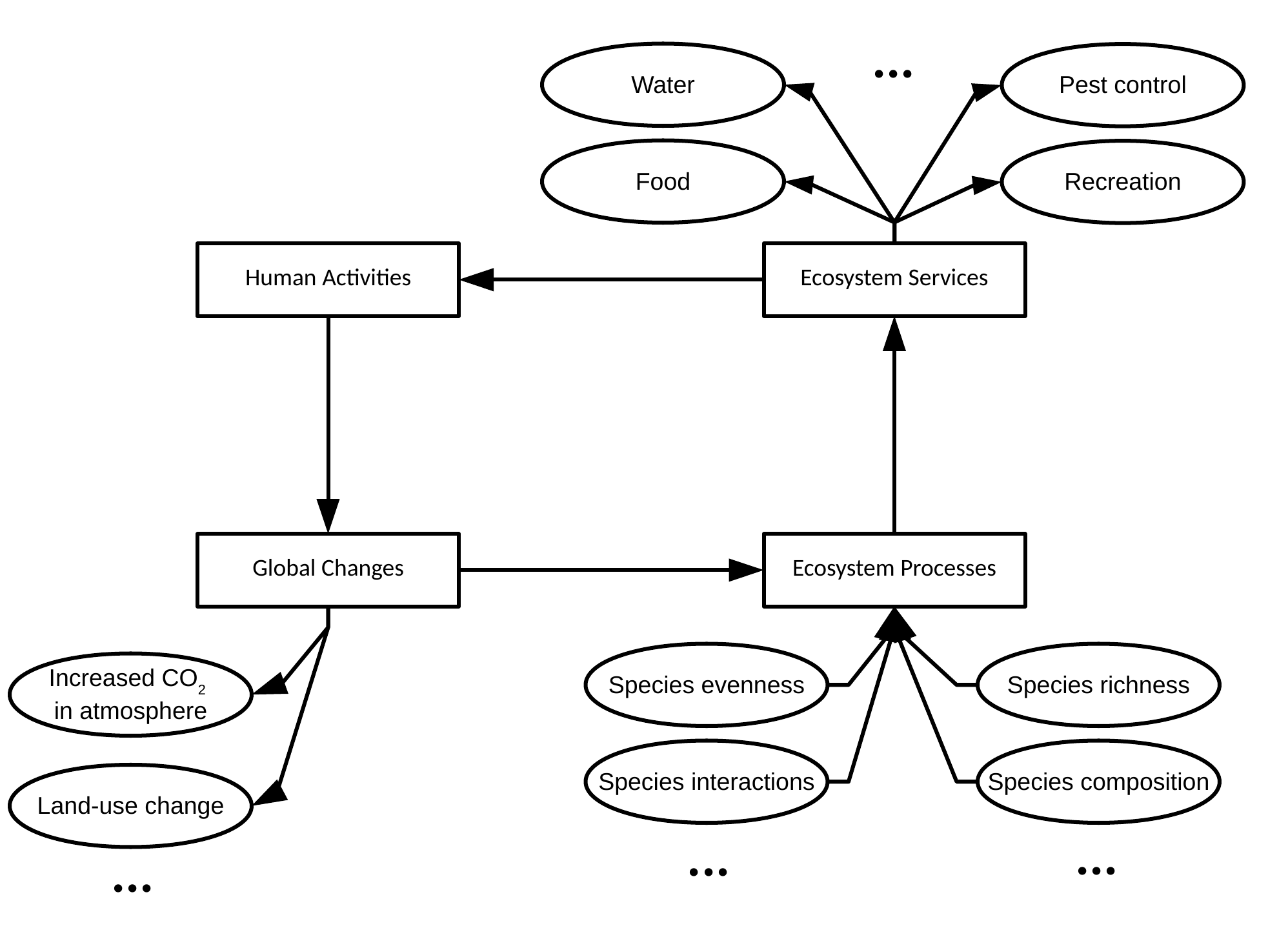}
\caption{Relationship between global changes and biodiversity (modified from \cite{Chapin2000}).\label{global_change}}
\end{center}
\end{figure*}

A major effort to address the problem was started in 1992 during the Earth Summit in Rio de Janeiro with the signature of the Convention on Biological Diversity (CBD) \cite{CBD1992}, a legally-binding international treaty. Its main objectives are the conservation of biodiversity, including ecosystems, species, and genetic resources, and their sustainable and fair use. Countries are required to elaborate and execute a strategy for biodiversity conservation, known as a {\em National Biodiversity Strategy and Action Plan} (NBSAP), and to put in place mechanisms to monitor and assess the implementation of this strategy. They should periodically report their progress on implementing their NBSAPs. Brazil, for instance, has advanced considerably in the creation of protected areas but recent expansion of agribusiness, mining, and hydroelectric power projects have threatened these advances \cite{Ferreira2014}. 
The {\em Strategic Plan for Biodiversity 2011-2020} defines actions to be taken by countries to achieve a set of twenty targets by 2020, known as the {\em Aichi Biodiversity Targets}. It should be observed that the United Nations General Assembly declared 2011-2020 the {\em United Nations Decade on Biodiversity}. In 2012, Intergovernmental Platform on Biodiversity and Ecosystem Services (IPBES) was created to allow for closer cooperation between scientists and policy makers on assessing the status of biodiversity and ecosystem services and their relationship.

To meet targets on biodiversity conservation, Balmford et al. \cite{Balmford2005} observe that it is essential to make indicators and knowledge openly available to decision-makers in ways that they can effectively use them. The {\em Group on Earth Observations Biodiversity Observation Network} (GEO	BON) proposed a set of twenty-two {\em Essential Biodiversity Variables} (EBVs) \cite{Pereira2013,Proenca2017GlobalVariables} that should allow for monitoring and evaluating biodiversity change. The development and deployment of mechanisms to produce these indicators depend on having access to trustworthy data from field surveys and automated sensors, biological collections, molecular data, and historic academic literature. Peterson \cite{Peterson2017}, however, shows that there are information gaps across thematic and geographical areas, suggesting that there should be funding and training for institutions and personnel working on biodiversity analysis to allow for evaluating EBVs globally. The transformation of raw data into synthesized data that is fit for use requires going through many refinement steps. One should assess its quality \cite{Chapman2005} by evaluating its taxonomic, geographical, and temporal accuracy. In many cases, the geographic coverage of the data is limited requiring, for instance, the use of species' distribution models \cite{Phillips2006} to estimate the likelihood of a particular species to occur in some geographic region. The methodologies and techniques used to manage and analyze this data comprise an area often called {\em biodiversity informatics} (or {\em e-Biodiversity}) \cite{Bisby2000,Soberon2004,Guralnick2009,Hardisty2013,hobern_global_2013,LaSalle2016}. Guralnick and Hill \cite{Guralnick2009}, for instance, propose the concept of a {\em global biodiversity map} that would record global patterns of biodiversity and how they change over time,  derived various data sources such as remote sensing, biodiversity literature, biological collections, and DNA sequence databases. On top of such map, various analyses, such as species richness and distributions, can be periodically updated using most recent data. Hardisty et al. \cite{Hardisty2013} recently gathered requirements for biodiversity informatics systems. They include, for instance, an aggregator for taxonomic names that integrates the distinct existing checklists; the use of persistent identifiers to reference datasets, scientific workflows, and scientists; use of linked data and ontologies to enable integrating data on different aspects of biodiversity; advancing data quality and fitness-for-use evaluation methodologies.

\begin{figure*}[h]
\begin{center}
\includegraphics[width=0.87\textwidth]{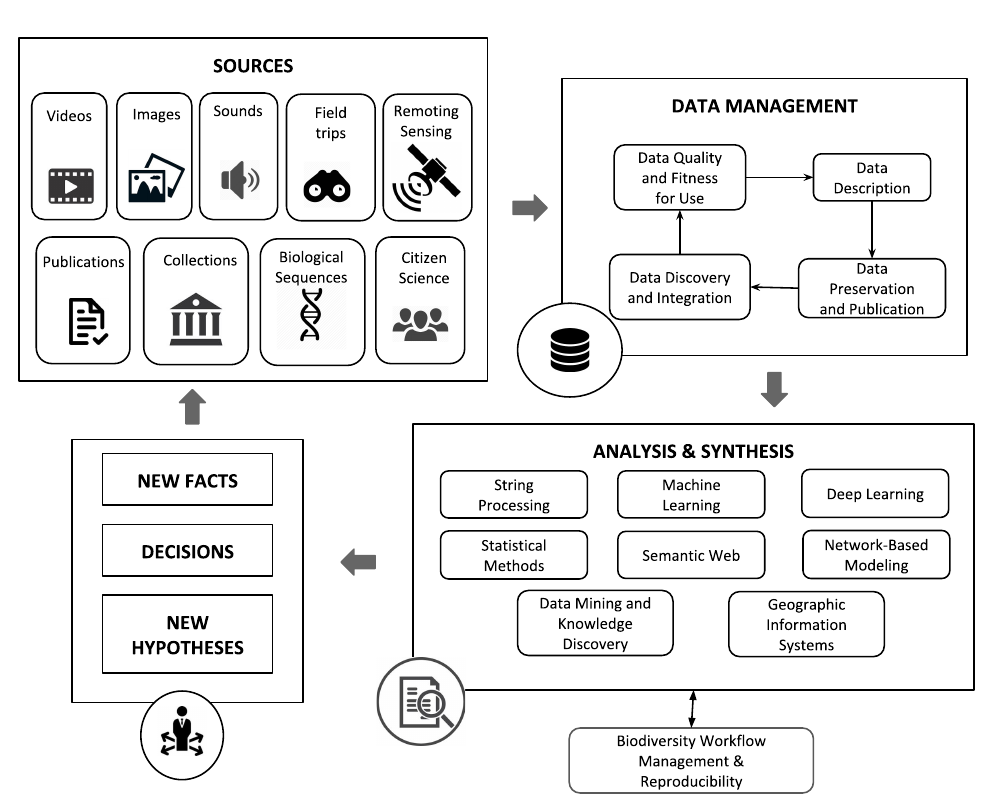}
\caption{Biodiversity Life Cycle.\label{biodiversity_life_cycle}}
\end{center}
\end{figure*}

In this survey, we give an overview of this research area covering its main concepts, practices, and some of the existing challenges. According to the approach proposed in \cite{Michener2012a}, biodiversity data follows a life cycle consisting of the following stages: planning, collection, certification, description, preservation, discovery, integration, and analysis. It is worth noting that after the analysis activity, new biodiversity data management cycles may be triggered as a result. In this survey, we group these steps in two main stages: data management and analysis and synthesis. Such steps are illustrated in Figure \ref{biodiversity_life_cycle}. Researchers, whether producers or consumers of biodiversity data will likely perform activities related to at least one of these steps. The remainder of this article addresses each stage of the life cycle of biodiversity data, describing their methodologies, tools, recommendations, and challenges.

% --------------
% --- Figure ---
% --------------

In Table \ref{tab:title}, we provide an overview of various tools and databases surveyed along with which steps of the e-Biodiversity life cycle they approach. In Table \ref{tab:computational}, we describe computational techniques used in each step of the e-Biodiversity life cycle that will be explored in this work.

\begin{table}[tbp]
\caption{A selection of e-Biodiversity databases and tools classified according to target life-cycle step: data planning and collection (DC), data quality and fitness-for-use (DQ), data description (DD), data preservation and publication (DP), data discovery and integration (DI), and computational modeling and data analysis (CM).\label{tab:title}} 
\begin{center}
%\footnotesize
{
\small \begin{tabular}{l|l|c|c|c|c|c|c}
Tool or database name & Reference  & DC & DQ & DD & DP & DI & CM \\\hline \hline
DMPTool &\cite{Strasser2014}  & $\checkmark$ & & & & & \\\hline
Morpho&\cite{Higgins2002}  & $\checkmark$ & $\checkmark$ & $\checkmark$ & $\checkmark$ &  & \\\hline
Metacat&\cite{Berkley2001}  & $\checkmark$ & $\checkmark$ & $\checkmark$ & $\checkmark$ & $\checkmark$ & \\\hline
Catalogue of Life&\cite{roskov2013species}  & $\checkmark$ & $\checkmark$ & $\checkmark$ & $\checkmark$ & $\checkmark$ & \\\hline
BHL&\cite{Gwinn2009}  &  $\checkmark$ &  $\checkmark$ &  $\checkmark$ & $\checkmark$ & $\checkmark$ & \\
\hline
eBird&\cite{Sullivan2014}  & $\checkmark$ & $\checkmark$ & $\checkmark$ & $\checkmark$ & $\checkmark$ & \\\hline
eMammal&\cite{He2016}  & $\checkmark$ & $\checkmark$ & $\checkmark$ & $\checkmark$ & $\checkmark$ & \\\hline
Brazilian Flora 2020&\cite{Forzza2012}  & $\checkmark$ & $\checkmark$ & $\checkmark$ & $\checkmark$ & $\checkmark$ & \\\hline
BRAHMS&\cite{Filer2013BRAHMSCourse}  & $\checkmark$ & $\checkmark$ & $\checkmark$ & $\checkmark$ & $\checkmark$ & \\\hline
Jabot&\cite{Silva2017}  & $\checkmark$ & $\checkmark$ & $\checkmark$ & $\checkmark$ & $\checkmark$ & \\\hline
MorphoBank&\cite{OLeary2011MorphoBank:cloud}  & $\checkmark$ & $\checkmark$ & $\checkmark$ & $\checkmark$ & $\checkmark$ &  \\\hline
BaMBa&\cite{Meirelles2015b}  & $\checkmark$ & $\checkmark$ & $\checkmark$ & $\checkmark$ & $\checkmark$ & \\\hline
Atlas of Living Australia&\cite{Belbin2016} & $\checkmark$ & $\checkmark$ & $\checkmark$ & $\checkmark$ & $\checkmark$ & $\checkmark$ \\\hline
speciesLink&\cite{Canhos2015} & $\checkmark$ & $\checkmark$ & $\checkmark$ & $\checkmark$ & $\checkmark$ & $\checkmark$ \\\hline
GenBank&\cite{Benson2013} & $\checkmark$ & $\checkmark$ & $\checkmark$ & $\checkmark$ & $\checkmark$ & $\checkmark$ \\\hline
MG-RAST&\cite{Meyer2008} & $\checkmark$ & $\checkmark$ & $\checkmark$ & $\checkmark$ & $\checkmark$ & $\checkmark$ \\\hline
BOLD&\cite{Ratnasingham2007} & $\checkmark$ & $\checkmark$ & $\checkmark$ & $\checkmark$ & $\checkmark$ & $\checkmark$ \\\hline
iPlant&\cite{Goff2011}  & $\checkmark$ & $\checkmark$ & $\checkmark$ & $\checkmark$ & $\checkmark$ & $\checkmark$ \\\hline
VoSeq&\cite{Pena2012}  & $\checkmark$ & $\checkmark$ & $\checkmark$ & $\checkmark$ & $\checkmark$ & $\checkmark$ \\\hline
SISS-Geo&\cite{Chame2018}  & $\checkmark$ & $\checkmark$ & $\checkmark$ & $\checkmark$ & $\checkmark$ & $\checkmark$ \\\hline
BioGeomancer&\cite{Guralnick2006} & & $\checkmark$ & &  & & \\\hline
Taxamatch&\cite{Rees2014} & & $\checkmark$ & &  & & \\\hline
Geospatial Data Quality&\cite{Otegui2016}  & & $\checkmark$ & & & & \\\hline
TNRS&\cite{Boyle2013a}  & & $\checkmark$ & & & & \\\hline
Kurator&\cite{Morris2018} & & $\checkmark$ & &  & & \\\hline
BioVel&\cite{Hardisty2016}  & & $\checkmark$ & & & $\checkmark$ & $\checkmark$ \\\hline
EU-Brazil OpenBio&\cite{Amaral2014}  & & $\checkmark$ & & & $\checkmark$ & $\checkmark$ \\\hline
Model-R&\cite{Sanchez-Tapia2018}  & & $\checkmark$ & & & $\checkmark$ & $\checkmark$ \\\hline
GGBN&\cite{Droege2014}  & & $\checkmark$ & $\checkmark$ &  & $\checkmark$ &  \\\hline
BioCollections&\cite{Sharma2018TheDatabase}  &  & $\checkmark$ & $\checkmark$ &  & $\checkmark$ &  \\\hline
GBIF&\cite{Edwards2004}  & & $\checkmark$ & $\checkmark$ & $\checkmark$ & $\checkmark$ & \\\hline
DataONE&\cite{Michener2012}                                              &              & $\checkmark$ & $\checkmark$ & $\checkmark$ & $\checkmark$ & \\\hline
OBIS & \cite{Grassle2000}  & & $\checkmark$ & $\checkmark$ & $\checkmark$ & $\checkmark$ &  \\\hline
SiBBr&\cite{BaringoFonseca2017}  & & $\checkmark$ & $\checkmark$ & $\checkmark$ & $\checkmark$ & \\\hline
IPT&\cite{Robertson2014}  & & & $\checkmark$ & $\checkmark$ & &\\\hline
Scratchpads&\cite{Smith2011}  &  & & & $\checkmark$ & $\checkmark$ & $\checkmark$ \\\hline
WholeTale&\cite{Brinckman2018}  & & & & & $\checkmark$ & $\checkmark$ \\\hline
Maxent&\cite{Phillips2006}  & & & & & & $\checkmark$ \\\hline
ConsNet&\cite{Ciarleglio2009}  & & & & & & $\checkmark$ \\\hline
OpenModeller&\cite{SouzaMunoz2009}  & & & & & & $\checkmark$ \\\hline
SPAdes&\cite{Bankevich2012}  & & & & & & $\checkmark$ \\\hline
SAHM&\cite{Morisette2013}  & & & & & & $\checkmark$ \\\hline
HipMer&\cite{Georganas2015}  & & & & & & $\checkmark$ \\\hline
SUPER-FOCUS&\cite{Silva2016}  & & & & & & $\checkmark$ \\
\end{tabular}}
\end{center}
\end{table} 

\begin{table}[htbp]
\begin{center}
\footnotesize
\caption{Computational techniques used in the e-Biodiversity life cycle: string processing (SP), metadata management (MD), conceptual modeling (CM), semantic web (SW), deep learning (DL), machine learning (ML), statistics (ST), geographical information systems (GS), graph theory (GT), parallel and distributed computing (PD), web services (WS). \label{tab:computational}} 
{
\small \begin{tabular}{l|c|c|c|c|c|c|c|c|c}
e-Biodiversity life cycle step          & SP & MD & CM & SW & DL & ML & ST & GS & GT \\
\hline \hline
Data Planning and Collection     &    &    &    &    &  $\checkmark$ &   $\checkmark$  &    &    &    \\
\hline
Data Quality & $\checkmark$   &    &    &    &  $\checkmark$ &  $\checkmark$  &    &  $\checkmark$  &    \\
\hline
Data Description                 &    &  $\checkmark$ &    &    &   &    &    &    &    \\
\hline
Data Publication &    &   & $\checkmark$ &    &   &    &    &    &    \\
\hline
Data Discovery and Integration &    &   &  &  $\checkmark$  &   &    &    &    &    \\
\hline
Ecological Niche Modeling &    &   &  &    &  $\checkmark$ &  $\checkmark$  &  $\checkmark$  & $\checkmark$   &    \\
\hline
Biodiversity Networks &    &   &  &    &   &    &    &    &  $\checkmark$  \\
\hline
Biodiversity Genomics & $\checkmark$    &   &  &    &   &  $\checkmark$   &  $\checkmark$   &    & $\checkmark$    \\
\hline
Wildlife Health Analysis &    &   &  &    &  $\checkmark$ &  $\checkmark$  &  $\checkmark$  & $\checkmark$   &    \\
\end{tabular}}
\end{center}
\end{table}

Throughout this work, we use definitions from the \textit{International Code of Nomenclature for algae, fungi and plants} (ICN) \cite{McNeill2012}. This document outlines a set of rules and guidelines for scientifically naming and grouping plants, fungi, and algae, consisting of a universally adopted reference by the botanical scientific community. Nomenclature best-practices for other groups of organisms are governed by other (though similar) documents.

Within the domain of biology, taxonomy is, in a general sense, the science of classification of organisms. Organisms are classified according to their shared characteristics and grouped at distinct levels of specificity (or \textit{taxonomic ranks}) using a hierarchical system, in which groups that are more specific are nested within broader ones. A taxonomic rank refers to the level of the taxonomic hierarchy at which a group of organisms is defined.  The most relevant ranks adopted in botany (in descending hierarchical order) are \textit{Kingdom}, \textit{Phylum} (or \textit{Division}), \textit{Class}, \textit{Order}, \textit{Family}, \textit{Genus}, \textit{Species}.
The taxonomic resolution of a biological sample is the rank of the most specific taxonomic determination that has been assigned to it. For instance, if a sample has been determined up to the level of \textit{species}, this rank is also its taxonomic resolution. As taxa relate to each other in a tree-like hierarchical structure (with each child taxon having exactly one parent, while a parent taxon can have one or more children), taxonomic identities of a specimen at ranks higher than its resolution can be directly determined. Although this term is not included in the ICN document, we use this definition throughout this text. A taxon is a taxonomic group of organisms at the level of any rank, which are considered by professional taxonomists to form a \textit{taxonomic unit}.  Species is one of the taxonomic ranks in which organisms can be classified. It is regarded to be a basic unit of taxonomic classification, although organisms can be further classified in lower-hierarchy taxonomic ranks (\textit{i.e.}, infraspecific ranks). Differently from other ranks, the name of a species is composed using a binomial nomenclature system, composed of the name of the genus followed by a \textit{specific epithet}, \textit{e.g.} \textit{Caryocar brasiliense}, \textit{Myrcia guianensis}, or \textit{Solanum lycocarpum}. 

When botanists sample organisms in the field, they either collect part of the organism (\textit{e.g.} a branch of a tree), the entire organism, or multiple individuals of the same type. Any of these collected biological materials is an evidence of the existence of a particular organism at some place and time and should be properly deposited in a biological collection for being preserved as a reference. A specimen is defined as one such evidence and refers to a punctual observation of a single kind of organism. Although a specimen could be classified by a taxonomist as being a representative of a given species, this is not a requirement for it to be included in scientific collections. Although taxonomists classify specimens in a best effort manner (the most taxonomically precise as possible), sometimes only higher ranks can be determined. The highest taxonomic rank at which the specimen could be identified is known as its \textit{taxonomic resolution}. After properly deposited in a biological collection, each record receives a taxonomic identification that assigns the individual to a \textit{taxon}.

Physical specimens stored in biological collections (also referred to as \textit{vouchers}) are often associated with complementary information, either annotated by the responsible collectors during the collecting act; or annotated at later stages, after the specimen is deposited in the collection \cite{Chapman2005}. Information from the collection event include the \textit{date, time}, and the \textit{geographic location} where the specimen was collected; the names of the \textit{collectors} who were involved in the collection event; and eventual \textit{field notes} describing contextual remarks, such as weather conditions, habitat features, or the sampling method used. Another crucial piece of information is the \textit{taxonomic identity} of the specimen, which can be determined by the collectors themselves or by professional taxonomists once the biological material is incorporated to the collection (although some materials eventually remain unidentified).

The taxonomic identity of a specimen includes not only the taxon name assigned to the sample, but also its nomenclatural status and authorship, the name of the person who has provided the identification, and in some cases, information regarding the certainty of identification.
As the taxonomic identity of a specimen can be re-evaluated by specialists several times after the first determination (though it requires that the investigator has access to the physical specimen), a history of determinations for specimens is usually stored in a collection. Vouchered specimens, together with their associated data, is what scientifically testifies a punctual observation of a species by a collector, at some location and at some point in time, and is thus referred to as a \textit{species occurrence}. 

e-Biodiversity, or mainly known as ``Biodiversity Informatics'', relates to the use of information technology (IT) to support the needs of understanding biodiversity, by organizing knowledge about individual biological organisms and the ecological systems they form. Over time, biodiversity informatics will deliver an increasingly interconnected digital resource supporting scientific research of the natural world \cite{hobern_global_2013}. The application of IT to support biodiversity research starts during the 1960s, tackling mainly two different aspects: handling scientific collections in museums and botanic gardens, and providing a fundamental tool for the development of the Numerical Taxonomy \cite{Perring1963,Yochelson1966,Silva1966,Smith1966,Sokal1966,MacDonald1966,Sokal1966a,Squires1966,Rogers1967,Crovello1967,Sarasan1981}. In the 1970's the use of computer to produce identification keys was explored \cite{Hall1970}, along with the expansion of their use in the management of herbarium records \cite{Beschel1970,Vitt1977,Wetmore1979}. In the same decade, we also have the initiative of the Flora of North America Program to ``create a computer data bank of taxonomic information about the vascular plants of North America north of Mexico.'' \cite{Krauss1973}. The decade of the 1980's is considered the ``golden era'' of the biodiversity informatics, characterized by a significant intellectual and bibliographic productivity and the sharing of ideas in scientific events. As result, iconic publications come to light, such as \cite{Allkin1984,Abbott1985}. 

In 1985, the foundations of a key organization in the development of the biodiversity informatics were launched: The Taxonomic Database Working Group (TDWG). The first meeting of TDWG was held at the Conservatoire et Jardin Botaniques, in Geneva, from 28th to 30th September 1985. In the first meetings, it became clear the need for elements of standardization for taxonomic databases and means for data exchange \cite{TDWG2017,Bisby1992}. In the same decade, we have witnessed the emergence of the first standards for data transfer \cite{IUCN-BGCS1987,White1992,Bisby1992,Conn1996} and the establishment of the first institutional databases and information systems, such as ``TROPICOS, run by the Missouri Botanical Garden; ILDIS, an international system covering legumes; BONAP, the Biota of North America Program; and CITES, Cactaceae list run at the Royal Botanic Gardens Kew. For microorganisms the impetus seems to have come from the culture collections and information industry; both BIOSIS and DSM ({\em Deutsche Sammlung von Microorganismen und Zellkulturen}) have established master lists for bacteria and others are available for yeasts and certain fungi'' \cite{Bisby1993}. 

At the end of the 1980s and beginning of the 1990s, the efforts for modeling taxonomic databases under the nomenclature rules and the dynamic of the classification became the concern of some researchers \cite{Beach1993,Pankhurst1993,Allkin1988,Allkin1992,Hine1991}. It is important to notice that it was during this period that small computers became more popular and affordable, added to the availability of database software and high-level programming languages, which made possible the creation of experimental databases dedicated to floristic and monographic data, as also morphological and chemical data, among others \cite{Allkin1984}. In September 1990, an interdisciplinary workshop, organized by University of California and funded by NSF, united different professionals to discuss the benefits of modern computer techniques - expert workstation for systematics, identification, phylogenetic trees, databases, and geographical information systems -- for systematic biology \cite{Fortuner1993}. In October 1990, a symposium called ``Designs for a Global Plant Species Database'', held in Delphi, Greece, focused in discussing different approaches for ``creating and operating a global plant species information system -- a data system that would provide international access to data accumulated on all of the world's plants.'' \cite{Bisby1993}.

In 1992, the CBD, in its article 7, item D, says: ``Maintain and organize, by any mechanism data, derived from identification and monitoring activities pursuant to subparagraphs (a), (b) and (c) above.'' In the same convention, the article 17 was dedicated to the ``Exchange of Information'', evocating the parties to facilitate the exchange of information of ``technical, scientific and socio-economic research, as well as information on training and surveying programmes, specialized knowledge, indigenous and traditional knowledge as such and in combination with the technologies referred to in Article 16, paragraph 1. It shall also, where feasible, include repatriation of information'' \cite{CBD1992}. This convention becomes the cornerstone of the Biodiversity Informatics area, coining its name as the result of the affiliation between five agencies, forming the ``Canadian Biodiversity Informatics Consortium'', to implement the CBD in Canada \cite{BGBM2010}. In 1999, a new milestone was set with the findings of the final report of the OECD Megascience Forum - Working Group on Biological Informatics, which understood that ``An international mechanism is needed to make biodiversity data and information accessible worldwide'', and ``recommends that the governments of OECD countries establish and support a distributed system of interlinked and interoperable modules (databases, software and networking tools, search engines, analytical algorithms, etc.) that together will form a Global Biodiversity Information Facility (GBIF)'' \cite{BGBM2010}. In September 2000 ``Bioinformatics for Biodiversity'' reached notoriety becoming cover of a special issue of the Science Magazine, consolidating itself as an area of research \cite{Sugden2000}. In 2004, the Global Biodiversity Information Facility (GBIF), an Internet-accessible interoperable network of biodiversity databases and information technology tools - went online with a prototype data portal\footnote{\url{http://www.gbif.org}} for simultaneously accessing data from the world's natural history collections, herbaria, culture collections, and observational databases \cite{Edwards2004}. In July 2012, around 100 experts from a wide variety of disciplines gathered in Copenhagen to the Global Biodiversity Informatics Conference (GBIC). The conference sets out a framework - The Global Biodiversity Informatics Outlook, launched in October 2013, to harness the immense power of information technology and an open data culture to gather unprecedented evidence about biodiversity and to inform better decisions \cite{hobern_global_2013}.

The remaining of this article is organized as follows.  In section \ref{sec:dm}, we describe the main steps of managing biodiversity data, including the tasks involved in planning and collecting biodiversity data (subsection \ref{sec:dpc}), data quality and fitness-for-use issues in biodiversity (subsection \ref{sec:dqffu}), the main metadata standards and tools for biodiversity (subsection \ref{sec:metadata}), the standards, tools, and systems that support publishing and preserving biodiversity data (subsection \ref{sec:dpp}), and techniques used to integrate biodiversity data from different sources and for discovering it (subsection \ref{sec:ddi}). In section \ref{sec:cmda}, we discuss existing tools for analysis and synthesis of biodiversity data, including computational models and areas of application. Finally, in section \ref{sec:conclusion}, we conclude the survey by presenting some of the current challenges of managing and analyzing biodiversity data.

% **********************************
% ********* DATA PLANNING **********
% **********************************
\section{Data Management} \label{sec:dm}

From planning and collecting biodiversity data to making it fit-for-use there are many steps that need to be followed, including data planning and collection, data quality and fitness-for-use, data description, data preservation and publication, and data discovery and integration. These steps comprise a biodiversity data management life-cycle, Figure~\ref{biodiversity_life_cycle} (top-right), that we present in this section.

\subsection{Data Planning and Collection}
\label{sec:dpc}

The various steps of the biodiversity life cycle usually comprise a {\em Data Management Plan} (DMP) \cite{Michener2012a} of biodiversity research activities. Some research funding agencies in countries like the United States require the submission of a DMP in submissions to calls for research proposals. Many funding agencies require that proposals should include a {\em data management plan}. A DMP is usually comprised of: which data will be collected; which formats or standards will be used for this data; which metadata will be provided and in which standard or format; what are the policies for data usage and sharing; how data will be stored and how it will be preserved in the long-term; and how data management will be funded. The DMP Tool\footnote{\url{https://dmptool.org/}}, for instance, is an online tool that supports designing and implementing a data management plan.

Biodiversity is concerned with the variety of living organisms, which can be measured in many different ways and scales, from a record of an organism observed in a geographical location at a particular date (a {\em species occurrence}) \cite{Yesson2007} to the relative abundance of species in a water sample collected at a long-term ecological research site \cite{Michener2011}. Omics also present many opportunities for exploring biodiversity. Molecular data from environmental samples \cite{Wooley2010,Robbins2012}, for instance, can be analyzed in metagenomics studies to identify functional traits and the taxonomic classification of organisms present. Biodiversity data can be collected in various ways: biosensor networks, field expeditions, observations made by citizen scientists, among others. In the collection process, it is important to use unique identifiers for project, sampling event, sampling area, and protocol used \cite{Stocks2016}. These identifiers will later allow the data collected to be stored in biodiversity databases consistently. Whenever possible, the terms should follow a controlled vocabulary or ontology, such as the Biodiversity Collections Ontology (BCO) \cite{Walls2014}. Next, we list common sources of data that are used to describe and analyze biodiversity:
\begin{itemize}
\item{\bf Species Occurrences}. Species occurrences are one of the most frequently available types of data concerning biodiversity. The main attributes of a species occurrence are given by: a {\em taxon}, which is defined as a group of one or more populations of an organism or organisms that form a unit;  a location; and a date of occurrence. Species occurrence records originate from different sources.  In order to facilitate the management and improve the accessibility of such information, most institutions currently maintain it organized in digital spreadsheets or in relational database systems, while also keeping references to the physical specimens they refer to. Some institutions are even deploying efforts towards digitizing the physical specimens. Hardisty et al. \cite{Hardisty2013} observe that only about 10\% of natural history collections are digitized and that tools are required to accelerate the process.   
In addition to specimens from biological collections, human observations are another source of species occurrences records. These observations take place, for instance, during field expeditions or even through citizen science initiatives (eBird \cite{Sullivan2014}, iNaturalist\footnote{\url{http://www.inaturalist.org}}). In some cases, species are maintained in the culture of living organisms, as is the case of various collections of fungi and other organisms.
\item {\bf Species Checklists}. Surveys are often performed within a geographic region, such as a continent \cite{Ulloa2017}, a country, or a national park, to determine which species are present in them. These surveys usually result in a list of taxon names called a {\em species checklist}. They might also be restricted to a particular kingdom or biome. Forzza et al. \cite{Forzza2012}, for instance, describe how the Brazilian Flora List published in 2010 was assembled, which involved aggregating information about species vouchers from herbarium information systems and having taxonomists to review it. The {\em Catalogue of Life}\footnote{\url{http://www.catalogueoflife.org}} aggregates more than 100 species checklists and contains information of about 1.6 million species.
\item {\bf Sample-based and Observational Data}. Sample-based data is collected during events, which may be one-time or periodical, typically involves environmental data and has a wide range and diversity of measurements. They may involve, for instance, abiotic measurements and population surveys in different temporal and spatial scales in transects, grids and plots \cite{Magnusson2013}. A typical context in which they are collected are the Long-Term Ecological Research (LTER) projects \cite{Michener2011} projects. Because of the heterogeneity of ecological data \cite{Reichman2011}, there is not a controlled vocabulary that is widely used. Some initiatives in this direction include ontologies such as ENVO, OBOE, and BCO \cite{Walls2014}. The most common tools for publishing ecological data rely on metadata to describe tabular datasets that comprise them. Such metadata allows general information, such as dataset owner identification, geographic, temporal, and taxonomic coverages, to be recorded, facilitating their interpretation by users. Metadata also allows textually describing the meaning of each column of a tabular dataset. Later in this article, the Ecological Metadata Language \cite{Fegraus2005a}, a metadata standard for ecological datasets, will be described.
\item {\bf Molecular Data}.

The analysis of DNA, RNA and proteins have various applications to the study of biodiversity. The genomic sequences obtained directly from environmental samples containing communities of microorganisms, i.e. metagenomes \cite{Robbins2012}, for instance, provide important information to analyze their taxonomic and functional characteristics. Biological sequences can support taxonomists as well \cite{Tautz2003} in identifying species.  Taxonomists can also use small genomic or gene regions to assess biological diversity accross all domains of life. The Barcode of Life \cite{Ratnasingham2007,Stockle2008} project, for example, analyzes and standardizes small regions of genes to help in identifying species. Some systems, such as VoSeq \cite{Pena2012}, allow for connecting vouchers present in biological collections to DNA sequences present in genomic databases. Guralnick and Hill \cite{Guralnick2009} observe that diversity can be more precisely measured, when compared to simply counting the number of species, by how species are phylogenetically related. As examples, they assess conservation priority of North American birds using their phylogenetic distinctness and extinction risk and analyze the dispersal of the influenza A virus also using phylogenetic analysis.  
\item {\bf Academic Literature}. A vast amount of information about the biodiversity is available in the academic literature. Field expeditions syntheses are often available only in scientific papers. Data related to sampling, collection, and their analysis have often not been propagated to biodiversity databases. Some initiatives, such as the Biodiversity Heritage Library (BHL) \cite{Gwinn2009}, aim to use technologies, such as optical character recognition, to extract this information from scientific articles and make them available in public databases.
\item {\bf Images and Videos}. Field expeditions to conduct sampling often involve the production of images and videos that support the analysis of the studied sites. In the following sections, we describe Audubon Core\footnote{\url{https://terms.tdwg.org/wiki/Audubon_Core}}, a controlled vocabulary for describing multimedia resources associated with sampling and species occurrence data.
\item {\bf Remote Sensing}. According to Turner et al. \cite{Turner2003}, most remote-sensing instruments do not have enough resolution to gather information about organisms but there were advances that enabled some aspects of biodiversity to be observed, such as differentiating species assemblages and tree species \cite{Clark2005}. They also argue that, when instrument resolution is insufficient for direct observation, indirect methods can be applied to estimate species distributions and richness. Pettorelli et al. \cite{Pettorelli2016} observe that many EBVs could be derived from satellite remote sensing, which can provide global-scale regular monitoring. Some of these potential EBVs include, for instance, vegetation height and leaf area index. It is also observed that raw satellite data could be processed by scientific workflows, including tasks such as statistical analysis and classification algorithms, to generate EBVs. 
\end{itemize}

% *********************************************
% ***** DATA QUALITY AND FITNESS-FOR-USE ******
% *********************************************
\subsection{Data Quality and Fitness-for-Use}
\label{sec:dqffu}

Although biodiversity scientists have undoubtedly benefited from open access to massive volumes of species occurrence data from many biological collections, there are some caveats that must be accounted for, before using data for modeling. Data is not always adequate for investigating every aspect of natural systems,
using inadequate data for studying specific aspects of biological diversity can lead to erroneous or misleading results \cite{Chapman2005}, and investigators must be aware of the inherent limitations of their data before formulating their questions. The availability of detailed information is still very scarce for most known organisms. This scenario, referred to as the \textit{Wallacean Shortfall} \cite{Lomolino2004ConservationBiogeography}, is even more critical in megadiverse countries, which still remain largely unexplored for many regions and taxonomic groups \cite{Soberon2004}. The lack of sufficient data for threatened species is even more concerning, as designing effective programs for their conservation require knowledge on their geographic distribution and ecological requirements. This shortage of data, combined with the non-systematic sampling and insufficient quality limits the use of data from biological collections for many intended applications, many of which require an intensive amount of data to be available \cite{Guisan2007WhatCharacteristics}. Failing to account for the inherent limitations of such data while posing and investigating their hypotheses, researchers may obtain erroneous or misleading results, eventually impacting the success of management policies that rely on such information \cite{Chapman2005}.

A definition for data quality based on its \textit{fitness for the intended use} was first proposed in the context of geographical information systems \cite{Chrisman1984}, and became widely adopted by the biodiversity informatics community. According to this definition, quality is not an absolute attribute of a dataset but is rather given by its potential to provide users with valuable information, in specific contexts. Assessing quality attributes of data is a fundamental step for any applications that might use it, and requires that users previously delimit the purpose, scope, and requirements of their investigation. Data is considered to be of high quality if it is suitable for supporting a given investigation. Depending on the application, users might need to improve the fitness of the data they have in hand, which is part of the data quality management process. Loss of quality in biodiversity data can occur during multiple stages of its life cycle \cite{Chapman2005}, including the moment of the recording event, its preparation before it is incorporated in the collection, its documentation, digitalization, and storage. Soberón and Peterson \cite{Soberon2004} list common issues regarding biodiversity data quality. Specimens of biological collections, from which a considerable amount of species occurrence data is extracted, may have incorrect or outdated taxonomic identifications. Biological taxonomy is constantly changing to accommodate new knowledge about species. It is also possible that there are georeferencing errors due to annotation error or instrument inaccuracy. In old records, due to the unavailability of mechanisms for accurate assessment of location, it is common to find only textual descriptions about where a specimen was collected.

One important aspect that often limits the usability of primary data from biological collections concerns the way in which it is gathered in the field. In general, most species occurrence data composing biological collections derive from exploratory field expeditions, in which organisms are recorded in a non-systematic \textit{observational} fashion by different collectors, using different methods and at distinct circumstances (though records resulting from experimental studies are eventually incorporated in museums as well). As a result, the distribution of the sampling effort in such datasets is uneven and rarely quantified, leading to \textit{sampling biases}. Building models without accounting for biases in data has been observed to strongly impact their performance, leading to spurious results which can be misinterpreted and, ultimately, lead to wrong decisions. For instance, assessing patterns of species richness from species occurrence datasets has been shown to be particularly challenging due to geographical bias in data \cite{Hortal2007LimitationsIslands,Reddy2003GeographicalAfrica}, as higher diversities tend to be observed at more accessible sites due to higher sampling effort. As defined by \cite{Chrisman1991}, biases are uniform shifts in measured values, resulting from systematic errors that are introduced by some measurement system. They are expressed as unrealistic tendencies in data, and can usually be mitigated with the adoption of random sampling designs.
Sampling bias in biodiversity data can be classified into several distinct categories, depending on the aspect of data under investigation \cite{Daru2017}. 

Not all taxa are quantitatively represented in biological collections in the same proportions as they occur in natural systems leading to \textit{taxonomic bias}. Collection sites are not randomly selected in geographic space, nor they are all sampled to the same extent. As features of the landscape make some areas more accessible for collection activities than others \cite{Hijmans2008AssessingPotatoes}, \textit{geographic bias} arises as a consequence of non-uniform collecting effort in geographic space. Some regions that are more accessible being thoroughly sampled (such as areas near urban centers, roadsides, and margins of rivers); while others that are more inaccessible, such as rainforests, being only poorly or not sampled at all.
Geographic bias is also observed at broader scales. A compilation of the representativity of plants in GBIF by~\cite{Meyer2016MultidimensionalInformation} has shown that among the most representative countries and regions are the United States (mainly the west coast), Central America, countries in Europe (including the Nordic countries), Australia, Japan, and New Zealand. The patterns of the recording activities of collectors are not uniform over time. Instead, collectors often show preferences towards performing field work in periods when they can get more productive, have more financial resources, or can find more organisms of their interests, leading to \textit{temporal bias}.

One of the objectives of CBD is to establish a global knowledge network on taxonomy  \cite{Hardisty2013}. Taxonomic concepts \cite{Berendsohn1995} are often incorrectly modeled in biodiversity databases. Berendsohn \cite{Berendsohn1997} developed a conceptual database model for the International Organisation for Plant Information covering the different aspects and concepts that are present in taxonomy. Several tools can be used to reduce or eliminate species misidentification. For instance, official species catalogs are available online for taxon querying, such as the Catalog of Life, the World Register of Marine Species\footnote{\url{http://www.marinespecies.org/}}, and the Brazilian Flora Species List \cite{Forzza2012}. These can be used to support taxonomic data quality assessment of occurrence records. Most of these catalogs are also accessible via application programming interfaces (APIs) available via the web, allowing the automation of this type of assessment with scripts or applications.

Dalcin \cite{Dalcin2005} investigates data quality in taxonomic databases, proposing quality metrics and techniques for error prevention, detection, and correction. He explores which dimensions are present when evaluating taxonomic data quality. {\em Accuracy} would measure how correct and reliable the data is. {\em Believability} is defined as how true and credible data is. {\em Completeness} measures to what extent attributes or data are missing. {\em Consistency} is defined as the absence of contradictions in the data. {\em Flexibility} is the capacity of representation of data to change in order to accommodate new requirements. {\em Relevance} is given by how applicable and helpful data is to a particular task. {\em Timeliness} how up-to-date the data is. Some experiments are conducted on detecting spelling errors in scientific names, which may be caused by insertions, deletions, substitutions, transpositions, or combination of the previous. Two types of algorithms are applied, {\em phonetic algorithms}, such as Soundex \cite{Holmes2002}, are based on pronunciation similarity. {\em  String similarity algorithms}, such as Levenshtein distance \cite{Levenshtein1966}, were also evaluated. Levenshtein algorithm showed a high incidence of {\em false alarms}. Phonetic algorithms had lower execution time than string-similarity ones. More recently, Rees \cite{Rees2014} observes that taxonomic names can contain errors due to misspelling which can lead to failure in retrieving data. He proposes Taxamatch, a method for approximate matching of taxonomic names. It uses a modified version of the Damerau-Levenshtein Distance \cite{Wagner1975} algorithm for genus and species name matching and a phonetic algorithm for authority matching. Experiments showed that the method is able to identify close to 100\% of errors in taxon scientific names.  \cite{Hardisty2013} observes that there are studies about biodiversity that do not require naming organisms. For instance, metagenomic studies concentrate on analyzing samples to classify them according to functional traits identified through sequence alignment with genomic databases. For collections that have digital images of their specimens available, a promising approach is to use deep learning techniques \cite{Schmidhuber2015DeepOverview} to automate species identification \cite{Bonnet2018PlantLearning}.

Regarding georeferencing problems, Guralnick \cite{Guralnick2009} mentions the importance of determining the georeferencing uncertainty of occurrence records and its impact on the scale at which studies can be performed. Tools like BioGeomancer \cite{Guralnick2006} and Geolocate\footnote{\url{http://www.museum.tulane.edu/geolocate/}} try to infer what the geographic coordinates of an occurrence of species from a textual location description. Otegui and Guralnick \cite{Otegui2016} propose a web API that performs simple consistency checks in occurrence records, such as coordinates with zero value, disagreeing coordinates and country identification, and inverted coordinates.

Veiga et al. \cite{Veiga2017} propose a framework for biodiversity data quality assessment and management that allows for users to define their data quality requirements and when a particular dataset is fit-for-use in a standardized manner. {\em Data quality assessment} is given by the evaluation of fitness for use of a dataset for some application. {\em Data quality management} is defined as the process of improving the fitness-for-use of a dataset. The framework is given by three main components: DQ Needs, DQ Solutions, DQ Report. DQ Needs supports the definition of the intended use for a dataset, the respective data quality dimensions, acceptable criteria for data quality measurements in these dimensions; and activities to improve data quality. DQ Solutions describe mechanisms that support meeting the requirements defined in the DQ Needs component, such as tools that implement techniques to improve data quality measurements in some dimension. The DQ Report component describes the dataset that is being assessed and managed by the framework and assertions on this dataset describing measurements or amendments applied to it as specified in the other components. The authors envision a {\em Fitness for Use Backbone} that would implement these components and where participants could share their data quality requirements and tools. More recently, Morris et al. \cite{Morris2018} have extended Kurator \cite{Dou2012}, a library of data curation scientific workflows, to report data quality in terms of the data quality framework proposed by Veiga et al. \cite{Veiga2017}.

% ******************************
% ****** DATA DESCRIPTION ******
% ******************************
\subsection{Data Description}
\label{sec:metadata}

In the description step, metadata is produced to describe biodiversity data. This metadata is essential for users to interpret datasets they download. In this section, we describe the standards, practices, and recommendations for documenting and describing  biodiversity data. 

% *****************
% ****** EML ******
% *****************
\subsubsection{Ecological Metadata Language}

The Ecological Metadata Language (EML) \cite{Fegraus2005a} is a metadata standard originally developed for the description of ecological data. It is also used currently to describe datasets about species observations. The standard has several profiles with their respective fields that can be used to define the attributes of a dataset. A scientific description profile contains fields such as the creator, geographic coverage ({\em geographicCoverage}), temporal coverage ({\em temporalCoverage}),  taxonomic coverage ({\em taxonomicCoverage}), and sampling protocol ({\em sampling}) used. This profile is used to define attributes of the dataset as a whole.

The data representation profile, through the {\em dataTable} entity, allows for describing the attributes of a tabular dataset. One can define the data types of such attributes, such as dates and numerical values, as well as their constraints, such as minimum and maximum values. Used together, the scientific description and the data representation profiles can provide good quality documentation for a dataset, supporting users to interpret it in a meaningful way.
EML metadata is expressed with the XML language, illustrated in Listing 1.

\scriptsize
\begin{figure}[h]
\begin{lstlisting}[frame=single, style=mystyle,language=XML,caption={Part of EML metadata in XML.}]
<dataset>  
<title>Baseline assessment of mesophotic reefs of West South Atlantic seamounts based on 
water quality, microbial diversity, benthic cover and fish biomass data</title>
<individualName>
<givenName>Fabiano</givenName>
<surName>Thompson</surName>
</individualName>
<organizationName>Federal University of Rio de Janeiro</organizationName>
<positionName>Assistant Professor</positionName>
<onlineUrl>http://www.microbiologia.biologia.ufrj.br</onlineUrl>
<abstract><para>Seamounts are considered important ...</abstract>
<coverage>
<geographicCoverage><geographicDescription>Vitoria Trindade Chain</geographicDescription>
<boundingCoordinates><westBoundingCoordinate>-38.875</westBoundingCoordinate>
<eastBoundingCoordinate>-16.375</eastBoundingCoordinate>
<northBoundingCoordinate>-17.125</northBoundingCoordinate>
<southBoundingCoordinate>-21.375</southBoundingCoordinate>
</boundingCoordinates>
</geographicCoverage>
<temporalCoverage>
<rangeOfDates>
<beginDate><calendarDate>2009-03-13</calendarDate></beginDate>
<endDate><calendarDate>2009-03-22</calendarDate></endDate>
</rangeOfDates>
</temporalCoverage>
</coverage>
</dataset>
\end{lstlisting}
\end{figure}
\normalsize

Normally, biodiversity databases provide tools for editing and producing metadata in the EML standard in a more user-friendly way through a graphical interface. The data repository DataONE \cite{Michener2012}, for example, allows users to provide metadata through a graphical tool called Morpho \cite{Higgins2002}. The same repository has also a web interface called Metacat \cite{Berkley2001}, which allows for loading  tabular ecological data in free format documented with the EML standard. The EML standard is also used to describe datasets on species occurrences and sampling events, as will be described in the following section. 

% ******************************
% ***** DATA DISSEMINATION *****
% ******************************
\subsection{Data Preservation and Publication}
\label{sec:dpp}

In the preservation stage, biodiversity datasets are published in some database, such as DataONE and GBIF, where they will be available to the scientific community. These databases adopt practices of curation and management of the data aiming its preservation and availability in the long term. There are several possible procedures for publication, in this section standards and procedures for loading a set of data to a biodiversity database will be described. The publication workflow of the main current repositories will also be described.

For better management of biological collections \cite{Schindel2018TheCollections}, several systems for this purpose have been developed in the last decades. Among the common features in this category of software are the management of specimens, control of determination history, taxonomy, images associated with specimens, bibliographic references, curatorial management activities, user management, reports tracking the evolution of collections, printing labels in varied sizes and data quality. These systems also have specific controls for the different types of scientific collections, such as exsicates, bromeliads, DNA, photo library, fruits, fungi, woods, orchids, seeds, in vitro, living, zoological, conservation, and related materials. Among the main ones are the BRAHMS \cite{Filer2013BRAHMSCourse}, used in more than 80 countries, allowing work with botanical collections, Specify\footnote{\url{http://specifysoftware.org}}, which is used in more than 500 institutions worldwide for over 30 years, managing collections of flora and fauna; the Emu\footnote{\url{https://emu.axiell.com}} proprietary software for managing collections, including botanical ones; BG-Base\footnote{\url{http://www.bg-base.com}} also with more than 3 decades of use, widely used in botanical gardens and arboretums. Jabot \cite{Silva2017} is used since 2005 in the Rio de Janeiro Botanical Garden  and started to be shared in the model of cloud computing with more than forty herbaria in Brazil.

% -------------------
% --- Darwin Core ---
% -------------------
\subsubsection{Darwin Core}

Darwin Core \cite{Wieczorek2012} is a standard for representing and sharing biodiversity data. The standard consists of a list of terms related to biodiversity and their definitions. Darwin Core (DwC) discussions, evolution, and maintenance are conducted by TDWG (Biodiversity Information Standards), an association for the development and promotion of standards for recording and exchanging biodiversity data. DwC emerged as a term profile in the 1998 Species Analyst system developed by the University of Kansas for the management of biological collections. In 2002, it was adopted for the exchange of information in Mammal Networked Information System (MaNIS), a distributed system composed of several institutions that maintain biological collections of mammals. In 2009, the DwC standardization process was started, which was ratified in October of the same year at the TDWG annual meeting. DwC is based on the Dublin Core\footnote{\url{http://dublincore.org/}} standard, taking advantage of its terms for resource description such as {\em type}, {\em modified} and {\em license}, and complementing them with specific biodiversity terms, such as {\em catalogNumber} and {\em scientificName}. 

The DwC vocabulary terms are organized as follows. The {\em classes} indicate the categories or entities defined in the standard. Examples of classes are: {\em Event}, {\em Location} and {\em Taxon}. Each class has a set of {\em properties}, which are its attributes. For example, the {\em Location} class has attributes such as {\em country} and {\em decimalLatitude}. Finally, values can be assigned to properties, such as ``Chile'', $-33.61$ for the {\em country} and {\em decimalLatitude} properties, respectively. It is worth noting that it is recommended that, whenever possible, the values come from some controlled vocabulary, in the case of textual values, or some formatting standard, in case of numerical or temporal values. For example, species names from some recognized list of species, such as the Catalog of Life. Table \ref{tab:dwcocc} illustrates the representation of species occurrence data with DwC. These records come from a dataset published through the Brazilian Marine Biodiversity Database (BaMBa) \cite{Meirelles2015b} in GBIF\footnote{\url{http://www.gbif.org/dataset/1edcfe6d-da55-4d59-b30e-468cd21f8b0b}}.

% -------------
% --- Table ---
% -------------
\begin{center}

\begin{table}[h]
\caption{Species occurrences represented with DwC.}
\begin{tabular}{l l l l l}
\hline
{\em id} & {\em eventDate} & {\em decimalLatitude} & {\em decimalLongitude} & {\em scientificName}\\
\hline
6  & 2002-08-01 & -20.805828 & -37.761231 & {\em Alectis ciliaris} (Bloch, 1787) \\
118 & 2002-08-01 & -22.382222 & -37.587500 & {\em Balistes vetula} (Linnaeus, 1758) \\
141 & 2002-08-01 & -19.848744 & -38.134635 & {\em Caranx crysos} (Mitchill, 1815) \\
507 & 2002-08-01 & -20.525417 & -29.310350 & {\em Thunnus obesus} (Lowe, 1839)\\
\hline
\end{tabular}
\label{tab:dwcocc}
\end{table} 
\end{center}

Normally, a set of data in the DwC format is accompanied by metadata, which is defined in the EML (Ecological Metadata Language) standard \cite{Fegraus2005a}. In EML, fields such as the title, authors, geographic and temporal coverage of the dataset are found, which help users interpret datasets formatted in the DwC standard.

Like relational databases, datasets that follow the DwC format can contain multiple tables that are related through properties that are common to all of them. Such an organization allows, for example, sample data to be expressed also in this standard. Tables 2 and 3 illustrate this type of data organization to represent species sampling. Table 2 contains the sampling events, four in total. The {\em eventId} column contains an identifier for each event. The other columns describe the event date, latitude, and longitude, respectively. Table 3 contains counts of organisms for each event. The {\em eventId} column, describes which event in Table 1 the counts refer to. For example, the first two rows in the table refer to the event that has identifier 1, which is associated with a sampling performed on March 18, 2009.

% -------------
% --- Table ---
% -------------
\begin{center}
\begin{table}[h]
\caption{Sampling events represented with DwC.}  
\begin{tabular}{l l l l}
\hline
{\em eventId} & {\em eventDate} & {\em decimalLatitude} & {\em decimalLongitude} \\
\hline
\underline{{\bf 1}}  & 2009-03-18 & -20.51 & -38.07 \\
2 & 2009-03-18 & -20.57 & -34.80 \\
3 & 2011-02-11 & -20.50 & -25.35 \\
4 & 2011-02-11 & -20.47 & -24.80 \\
\hline
\end{tabular}
\label{tab:dwcsample}
\end{table} 
\end{center}

% -------------
% --- Table ---
% -------------
\begin{table}[htbp]
\caption{Species occurrences related to the events in Table \ref{tab:dwcsample}.} 
\begin{center}
\begin{tabular}{l l l}
\hline
{\em eventId} & {\em organismQuantity} & {\em scientificName} \\
\hline 
\underline{{\bf 1}}  & 2 & {\em Alectis ciliaris} (Bloch, 1787) \\
\underline{{\bf 1}} & 5 & {\em Balistes vetula} (Linnaeus, 1758) \\
2 &  1 & {\em Thunnus obesus} (Lowe, 1839) \\
$\ldots$ & $\ldots$ & $\ldots$ \\
\hline
\end{tabular}
\label{tab:dwcsample-occ}
\end{center}
\end{table} 

\begin{figure*}[h]
\begin{center}
\includegraphics[width=0.6\textwidth]{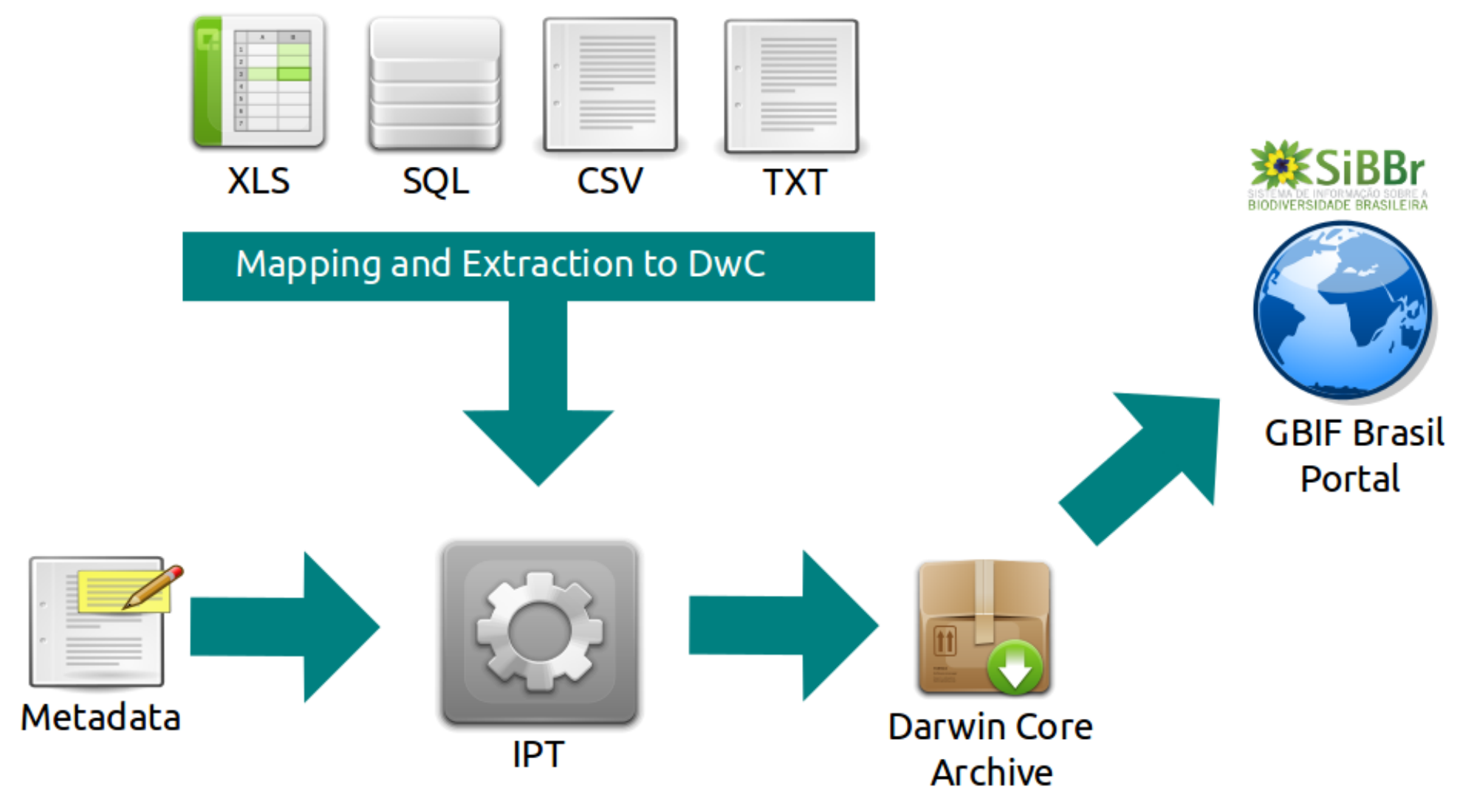}
\caption{Data publication using IPT.\label{ipt_diagram}}
\end{center}
\end{figure*}

Biodiversity datasets formatted with DwC can be published in global-scale biodiversity databases such as the GBIF \cite{Edwards2000}. GBIF acts as a central registry and aggregator for datasets published by its national and organizational nodes using the {\em Integrated Publishing Toolkit} (IPT) \cite{Robertson2014}. The publishing workflow includes the following steps: (i) mapping the internal representation of biodiversity data to DwC and extracting it; (ii) adding EML metadata describing the biodiversity data; (iii) packaging both EML metadata and DwC-formatted data to a {\em Darwin Core Archive} (DwC-A); (iv) GBIF and national biodiversity aggregators, such as the Brazilian Biodiversity Information System (SiBBr) \cite{Gadelha2014a,BaringoFonseca2017}, harvest DwC-A and ingest them into their databases. This process is illustrated in Figure \ref{ipt_diagram}.

% --------------
% --- Figure ---
% --------------

Many research groups have tabular data about biodiversity stored in various formats and do not have the resources to format them according to DwC.  Different approaches in ecology, coupled with distinct research traditions, both in their subdisciplines and in related fields, lead to the production of highly heterogeneous data. Such data can be, among others, counts of individuals, measures of environmental variables or representations of ecological processes. The terminologies used also vary according to the research line, as well as how to structure the data digitally \cite{Jones2006}. The EML metadata standard was also adopted for describing the ecological datasets. The datasets themselves, due to the heterogeneity, are published in the original format, through spreadsheets or textual files with values separated by commas. In these cases, Metacat \cite{Berkley2001} can support data publication and preservation. It is responsible for receiving, storing and disseminating datasets of, for example, Long-Term Ecological Research (LTER) \cite{Michener2011}. The Brazilian Marine Biodiversity Database (BaMBa) \cite{Meirelles2015b}, for instance, was developed to store large datasets from integrated holistic studies, including physico-chemical, microbiological, benthic and fish parameters. BaMBa is linked to SiBBr and has instances of both IPT and Metacat, making it possible to publish data using the workflows previously described.

The publication of data, and its consequent preservation, is a contribution to the scientific community as a whole. The data can be reused by other scientists who can explore them from other points of view. For the publisher of the data, benefits can also be observed. A recent study \cite{Piwowar2013} shows that articles that provide the data used in their analyses in public repositories tend to have a larger number of citations. Data publication and preservation can also be helpful when biological collections are lost due to disasters. In September 2, 2018, there was a catastrophic fire in the Brazilian National Museum, destroying the vast majority of approximately 20 million items in its collections comprising areas such as archeology, anthropology, zoology, and botany. Many destroyed items belonged to biological collections, including one on invertebrates. Through data publication on GBIF\footnote{\url{https://www.gbif.org/publisher/4205110f-3f0f-40d8-bd0f-2fa71bc827b5}}, the museum was able to preserve information about many specimens, 269,660 records were available in September 20, 2018, many of these containing images. 

\subsubsection{Other Standards}

Access to Biological Collection Data (ABCD) provides a common definition for content data from living collections, natural history collections, and observation datasets. It also offers a detailed treatment of provider rights and copyright statements. In many cases, it defines elements for both highly atomized and less structured data to encourage potential providers to take part in information networks even if their collection databases are less atomized or not normalized \cite{Guntsch2007}.   
Currently, ABCD is used to publish data, for instance, in GBIF.
Multi-media resources can provide reliable evidence for the occurrence of a taxon in a particular place and time, and there is a growing recognition that a biodiversity-related multimedia object could be used as a ``primary biodiversity record'' if the metadata associated with the object is available and has high quality. Therefore, Audubon Core  \cite{Morris2013a} standard came to fill the gap of a standard related to multimedia resources, as digital or physical artifacts that normally comprise more than text. These include photographs, artwork, drawings, sound, video, animations, and presentation materials, as well as interactive online media such as species identification tools.
The Audubon Core is a set of vocabularies designed to represent metadata for biodiversity multimedia resources and collections. These vocabularies aim to represent information that will help to determine whether a particular resource or collection will be fit for some particular biodiversity science application before acquiring the media. Among others, the vocabularies address such concerns as the management of the media and collections, descriptions of their content, their taxonomic, geographic, and temporal coverage, and the appropriate ways to retrieve, attribute and reproduce them \cite{Morris2013a}.
Plinian Core \cite{Riva2013,Pando2017} is a set of vocabulary terms that can be used to describe all kind of properties related to taxa and in its actual version, Plinian Core incorporates a number of elements already defined within standards in use, such as EML and DwC.

% ******************************************
% ***** DATA DISCOVERY AND INTEGRATION *****
% ******************************************
\subsection{Data Discovery and Integration}
\label{sec:ddi}

The search for data to perform biodiversity analysis and synthesis research is still a challenging task. The most recent developments have occurred with the emergence of databases that aggregate datasets at global and national scales such as GBIF \cite{Edwards2000}, DataONE \cite{Michener2012}, SiBBr \cite{Gadelha2014a} and BaMBa \cite{Meirelles2015b}. The use of metadata and data publishing standards allows institutions to map the internal representations of this information to a format that is clearly specified and can be consumed and processed automatically by machines. Biodiversity information aggregation databases allow datasets to be geographically, taxonomically, and temporally searched. Languages and data analysis environments, such as R and Python, already have packages and libraries that are integrated with the repositories and aggregators of biodiversity data. {\em rgbif}\footnote{\url{https://cran.r-project.org/web/packages/rgbif/}}, for example, is a package for R that allows searching and retrieving records directly from GBIF, with {\em pygbif}\footnote{\url{https://recology.info/2015/11/pygbif/}} being its analog for Python.

Often scientists need to combine data from different sources into integrative research. For example, physico-chemical data can be combined with metagenomic data to try to establish correlations that explain some ecosystem processes and implications such effectiveness of marine protected areas \cite{Bruce2012}, contribution of sea-mounts \cite{Meirelles2015}. The activity of combining data from different sources is called data integration and is one of the most active areas of research on scientific data management \cite{Ailamaki2010,Miller2018OpenIntegration}. Existing biodiversity databases have advanced by establishing standards for metadata, such as EML \cite{Fegraus2005a}, and for data such as DwC. However, these are limited to defining controlled vocabularies, consisting of standardized terms in each of the themes. A more sophisticated approach, involving not only the definition of terms but also the relationships between them and rules of inference, which are called ontologies, is the subject of the {\em Semantic Web} research area. Some initiatives in this direction in the area of biodiversity and ecology include ontologies such as the {\em Environment Ontology} (ENVO) and the {\em Biodiversity Collections Ontology} (BCO) \cite{Walls2014}. Ontologies allow cross-referencing of different domains ({\em Linked Data}) and semantic queries, providing a data integration tool considerably more powerful than the current ones.

% ****************************************************
% ***** COMPUTATIONAL MODELING AND DATA ANALYSIS *****
% ****************************************************
\section{Data Analysis and Synthesis}
\label{sec:cmda}

In this section, we present some examples where biodiversity data is analyzed along with the computational methods used. These main biological examples are related with ecological niche modeling, network science, biodiversity genomics, wildlife health monitoring. We also explore methods for interconnecting various  computational tasks, i.e. biodiversity workflow management, and for keeping track of data derivation in these workflows with the purpose of enabling analysis and synthesis reproducibility.

% ---------------------------------
% --- ECOLOGICAL NICHE MODELING ---
% ---------------------------------
\subsection{Ecological Niche Modeling}
\label{subsec:enm}

Ecological Niche Modeling (ENM) is used to predict the potential geographic distribution of a given species, based on environmental factors \cite{TownsendPeterson2011}. A niche-based model represents an approximation of the ecological (realized) niche of a species in the environmental dimensions analyzed \cite{Phillips2006,TownsendPeterson2011}. This model is made using a family of statistical tools to analyze the environmental information associated with the occurrence points (geographic coordinates), generating maps with an indication of geographic areas with the environmental suitability of the modeled species \cite{Elith2009,Gomes2018SpeciesData}.
Different studies include ENM with the objective of analyzing, for example, the potential distribution of invasive species (e.g. \cite{Peterson2003}), with indication of vulnerable areas, the distribution of species in scenarios of climate change (e.g. \cite{Siqueira2003,Thomas2004,Pearson2006,Araujo2008,Wiens2009,Araujo2012}, dissemination of infectious diseases (e.g. \cite{Costa2002}), and selecting conservation areas (e.g. \cite{Araujo2000a,Engler2004a,Chen2009,Pearson2010}). 
The concept of niche is defined, according to Chase and Leibold \cite{Chase2003}, as environmental conditions that meet the minimum requirements of a species so that its birth rate is higher than its mortality rate. There are three main factors that determine the niche of a species: abiotic (environmental) conditions, biotic conditions, such as species interactions, and dispersal capacity \cite{Soberon2010}. These are illustrated by the BAM diagram which depicts the biotic factors (B), the abiotic factors (A) and the mobility (M) \cite{TownsendPeterson2011}.

\begin{figure*}[h]
\centering\includegraphics[width=1.0\linewidth]{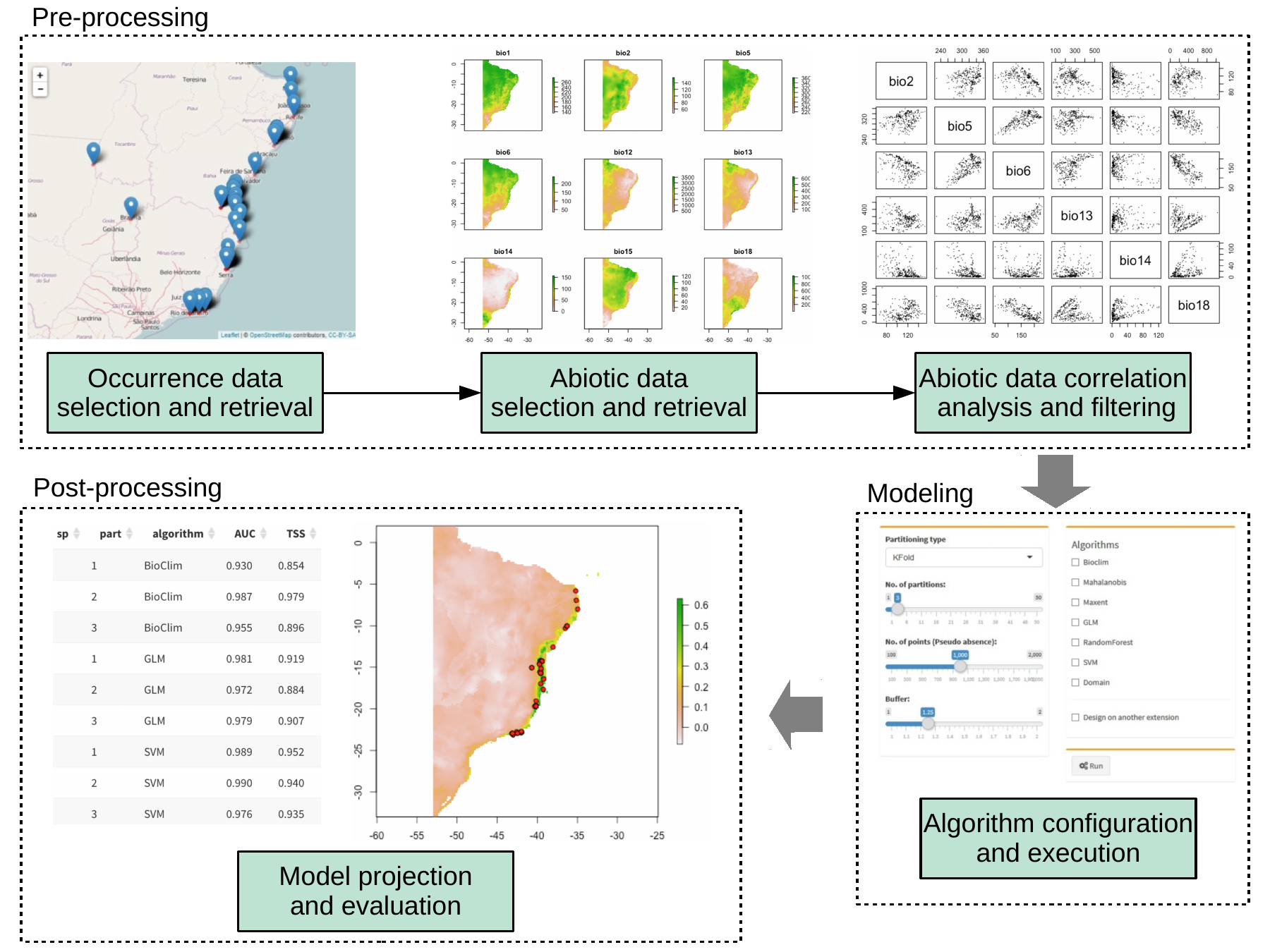}
\caption{Typical ENM steps comprising (1) pre-processing (occurrence data selection and retrieval, abiotic data selection and retrieval, and abiotic data correlation analysis and filtering), (2) modeling (algorithm configuration and execution), and (3) post-processing (model projection and evaluation).\label{model-r-web}}
\end{figure*}

The Ecological Niche Modeling involves three general steps: (1) Pre-processing, (2) Modeling and (3) Post-processing. In the pre-processing stage, after the acquisition of species occurrence records (biotic variables), in databases such as SpeciesLink \cite{Canhos2015} and GBIF \cite{Edwards2004}, the coordinates must be verified, to eliminate positional uncertainty and inconsistencies, such as inverted longitude and latitude. Abiotic variables can be downloaded, for example, in climatology databases such as Worldclim \cite{Hijmans2005} and CHELSA \cite{Karger2017ClimatologiesAreas}. The environmental layers are downloaded and converted so they can be used as input to model algorithms along with occurrence points. These datasets are usually in raster format, that is, a grid of two-dimensional cells, where each cell has a value. The resolution of the dataset is given by the size of a cell and the smaller the cell, the higher the resolution. In this step, we also verify the sample bias, using a) spatial filters - to remove points very close geographically, in order to select points with a minimum geographical distance between them \cite{Naimi2011,Boria2014,Varela2014}. This procedure aims to reduce the spatial bias effects of points. And b) environmental filters - to remove points with low variation in environmental layers, avoiding environmentally close points \cite{Loiselle2007,Varela2014}. This procedure aims to reduce the environmental autocorrelation of the data, and the use of clean data generates models with greater predictive power \cite{Calabrese2014StackingModels,Lahoz-Monfort2014ImperfectModels,Aiello-Lammens2015SpThin:Models}. The modeling step consists in the application of algorithms and the occurrence and biotic and abiotic data raised in the pre-processing phase, for the creation of the models. Various algorithms are used in ecological niche modeling, some based on machine learning or environmental envelopes. Machine learning algorithms include Maxent \cite{Phillips2006}, Genetic Algorithm for Rule-Set Production (GARP) \cite{Stockwell1999}, both require only presence points. Other algorithms include Generalized Linear Models (GLM) \cite{McCullagh1989}, Generalized Additive Models (GAM) \cite{Hastie1990}, and Random Forest  \cite{Breiman2001}. Post-processing is the evaluation stage of the generated model. The validation of a model is made from the comparison of the generated results, against distribution data of the species not used in the modeling process. Among the post-processing procedures, analyses are performed to increase reliability, or to reduce the uncertainty of the models generated by different algorithms \cite{Giannini2012}. A consensus model is generated from means of combined projection techniques, where high suitability areas coincide in most of the models generated for a given species \cite{Araujo2007}. The calculation of the area under the curve (AUC) is the most used validation test \cite{Giannini2012}, but a number of others are applied to evaluate the performance of the models, such as Kappa (Cohen's Kappa Statistic), True Skill Statistics (TSS), Lowest Presence Threshold (LPT) \cite{Allouche2006,Elith2006,Pearson2006a,Liu2011} and more complex strategies with a combination of maps, considering, for example, geographic barriers, deforested areas \cite{Anderson2003,Liu2005,Pearson2006a}, and multidimensional analyses \cite{Diniz-Filho2009}.
A framework for scalable and reproducible ecological niche modeling, Model-R \cite{Sanchez-Tapia2018} was developed with the objective of unifying pre-existing ecological niche modeling tools in a web interface that automates processing steps and modeling performance. This tool includes packages related to data pre-processing like retrieve and cleaning data (e.g. RJabot, a feature that allows you to search for and retrieve Jabot \cite{Silva2017} occurrence data), multi-projection tools that can be applied to different temporal and spatial datasets and post-processing tools linked to the generated models. The Model-R has seven algorithms implemented and available for modeling: BIOCLIM, Mahalanobis distance, Maxent, GLM, RandomForest, SVM, and DOMAIN. The algorithms, as well as the entire modeling process, can be parameterized using command line tools or through the web interface. Figure \ref{model-r-web} describes typical ENM steps. 

So far, ecological niche modeling has relied mostly on abiotic variables. A challenge is to incorporate biotic information as well, such as species interactions. According to \cite{TownsendPeterson2011}, these are difficult to gather since they often require surveys at local scales. One experiment using biotic information was performed by \cite{Heikkinen2007BioticMacro-scales} incorporating mutualism information to model four bird species, which improves the accuracy of prediction considerably.

% ----------------------------------------------------
% --- NETWORK SCIENCE APPLIED TO BIODIVERSITY DATA ---
% ----------------------------------------------------
\subsection{Network Science Applied to Biodiversity}

Network Science refers to a relatively new domain of scientific investigation, aiming to describe emergent properties and patterns from complex systems of interacting entities. Such relational systems are naturally represented as networks, in which interactions are represented as pairwise connections (\textit{links}) between entities (\textit{nodes}) and assume particular semantics depending on the nature of the modeled phenomenon. The rise of this field is strongly associated with recent advances of information technology, which provided scientists with novel tools for collecting, storing, and processing data from many knowledge domains more efficiently and in larger scales. Although a variety of networked systems in many disciplines had been studied long before that, technological advances allowed us to model real-world systems in much more detail, from large volumes data that are often public or easily accessible to the investigator. Network science \cite{Barabasi2016,Newman2010} has been applied to model networked systems in a variety of knowledge domains, including the Internet, scientific collaboration networks, and ecological networks just to cite a few \cite{Albert2002}. Given their relational structure, network models are formally represented as \textit{graphs}. Network modeling has been widely adopted in the context of biodiversity research, especially for investigating ecological and evolutionary aspects of ecosystems and natural communities.
Efforts toward this goal have led to the creation of the field of \textit{network ecology}, which has undergone a noticeable growth over the last few years \cite{Borrett2014}.
Network ecology has traditionally focused on describing general aspects of the entangled networks by which organisms interact.
As ecological interactions are regarded as key processes modeling ecosystems functioning and structure, unraveling their architecture and dynamics is essential for understanding a variety of ecosystem features, such as stability and energy flow.
Interaction networks can be broadly classified as \textit{food webs}, \textit{host-parasitoid webs}, or \textit{mutualistic webs} \cite{Ings2009}, being food webs the first ones described in literature, since two classical papers by \cite{Lindeman1942,Odum1956}. 

Besides ecological interactions, network thinking has also been applied for modeling other aspects of natural systems. Patterns of animal movement can be investigated in a structured way, for instance, by means of \textit{movement networks} \cite{Jacoby2016EmergingEcology}.
These networks represent geographical space as a set of discrete and interconnected locations, forming a mesh of possible routes through which animals (or groups of animals) travel.
Links between each pair of locations are weighted according to their geographical connectivity.
Animal movement is thus regarded as dynamic processes composed of sequences of discrete movement steps running through the network structure. As the spatial feature is key in this type of network, they are also referred to as \textit{spatial networks}~\cite{Bascompte2007NetworksEcology}.

Others have applied network science to investigate biogeographical patterns, such as species co-occurrence. The so-called \textit{co-occurrence networks} model species associations in terms of their geographical distributions, such that species which are often observed occurring together in the same set of localities are considered to be strongly connected to each other.
Similarly to other networked systems, co-occurrence networks are composed of a majority of the species holding co-occurrence links to very few others, whilst only a few species are connected to many others~\cite{Araujo2011}. Co-occurrence network analysis has been used for many applications in biodiversity studies, such as for selecting subsets of species to be used as surrogates for the characterization of biological communities~\cite{Tulloch2016DynamicSurrogates};
for assessing the resilience of biotic communities towards climate change~\cite{Araujo2011};
and for identifying modularity (clusters of overlapping species ranges) in biological communities from animal-location bipartite networks~\cite{Thebault2013IdentifyingMeasures}.

The social network analytics framework has also been applied in some biodiversity studies, though in most cases for modeling animal social behavior \cite{Faust2011AnimalNetworks}.
An alternative perspective is to look at communities of biodiversity data producers and consumers, in order to better understand the myriad of contexts in which data is collected, shared, and used.
Mapping data flow within the community of biodiversity informatics initiatives, for instance, could help to prioritize and to improve the coordination of collaborative actions, leading to more effective biodiversity data-based policies \cite{Bingham2017TheOpportunities}.
Further, important scientific communities and gaps could be identified and characterized by exploring collaborative paper authoring networks and scientific topic networks \cite{Borrett2014}.
Another interesting example of a collaboration network in biodiversity is given in \cite{Groom2014Herbarium18561932}, where a correspondence network of $19th$-$20th$ century botanists was structured from digitized data from the British Herbaria. Botanists composing this network corresponded with each other by exchanging specimens, a practice that has led to the formation of exchange clubs. Many aspects regarding the particular ways botanists used to work as well as the roles they assumed could be investigated with the aid of exchange networks.

Finally, a better understanding of the factors and processes influencing the composition of species occurrence datasets would be invaluable for improving data usability, especially for species distribution modeling \cite{Daru2017}. As biological collections are typically composed of an ensemble of opportunistic species occurrence records, each of which having been gathered in a particular context by a different collection team, their datasets do not necessarily reflect the biological diversity from the areas in which the collections are physically located. Rather, they best reflect the interests of their most active and relevant collectors, i.e. those who have contributed to the collection to larger extents.

% %%%%%%%%%%%%%%%%%%%%%%%%%%%%%
% %%% BIODIVERSITY GENOMICS %%%
% %%%%%%%%%%%%%%%%%%%%%%%%%%%%%
\subsection{Biodiversity Genomics}
 
The DNA is the universal code of all organism in the Earth, present from simple species, such as bacteria and viruses, to more complex organisms, such as the vertebrates. The complete DNA sequence of an organism defines its genome. Due to the advances in new sequencing technologies, called Next Generation Sequencing (NGS), unknown genomes have been sequenced, assembled and deposited in public data repositories of molecular data. This data is growing fast because of the decreased cost of NGS and increased capacity of computational infrastructures \cite{Stephens2015BigGenomical,Lee2018ExascaleBiology}. The advances in genomic information bring results in many application areas to society~\cite{Lewin2018EarthLife}, including the production of valuable bioproducts at industrial scale (biofuels, bioenergy, cellulose fibers, gum chemicals, oils and resins), biomonitoring of species (for instance, viruses in epidemiological surveillance \cite{Holmes2008EvolutionaryViruses}), new drugs (vaccine design \cite{He2015ComputationalEvaluation} and protein therapeutics \cite{Leader2008ProteinClassification}). 

Molecular approaches are becoming one the most relevant tools to support the taxonomist in species identification \cite{Hebert2003BiologicalBarcodes}. The community is looking for the genes or regions of the genome as DNA barcode candidates, such as COI (cytochrome C oxidase I), used to identify animals (mammals, insects, fishes); ITS (internal transcribed spacer) for fungi; ribosomal RNA (rRNA) -- subunit 16S for identify bacteria; matK (maturase K) and rbcL (ribulose-bisphosphate carboxylase) to identify plants. BOLD (Barcode of Life Data System)  \cite{Ratnasingham2007} provides an integrated bioinformatics platform that assists in the acquisition, storage, analysis and publication of DNA barcode records. It is developed and hosted by the International Barcode of Life project (iBOL), one of the largest biodiversity genomics initiative ever executed. Hundreds of biodiversity scientists, bioinformaticians and technologists from 25 nations are working together to construct a richly parameterized DNA barcode reference library that will be the foundation for a DNA-based identification system for all multi-cellular life~\cite{Page2008}.

In general, the macroscopic species are identified and cataloged by morphological aspects and receive a voucher identifier which contains information about this species and metadata (geographical location, collector, collection date, etc.). The Global Genome Biodiversity Network (GGBN)~\cite{Droege2014,Droege2016} data portal, for instance, stores information about vouchered collections of DNA or tissue samples. Museums and institutions are joining efforts to link their biological collection with genetic data as nucleotide sequences from particular genes. Genbank (from NCBI), DDBJ (DNA Databank of Japan), ENA (European Nucleotide Archive), which are part of the International Nucleotide Sequence Databases Collaboration (INSDC), are the most popular repositories for nucleotide sequences. Although the voucher tag is present in Genbank since 1998, it remained poorly used~\cite{Schoch2014FindingFungi}. Recently, the BioCollections database from NCBI connected specimen vouchers to sequence records in GenBank~\cite{Sharma2018TheDatabase}. 

Other interesting area comprised the recover of the DNA of the extinct species, denominated ancient DNA (or aDNA) which provides resources to understand the evolutionary process. The reconstruction of aDNA involves material derived from archaeological specimens, mummified tissues, preserved plant remains, and from other environments, such as permafrost and sediments \cite{Burrell2015TheSequencers}. Until now, most of the extinct species sequenced belong to mammalian megafauna and ancient humans \cite{Campbell2012}. The experimental difficulty and the challenges in this area are related to the quality of the material (which degrade with time) and contamination. 

Considering microscopic organisms, NGS opens a new world of capabilities reducing the need for culture isolation of microscopic species such as bacteria, viruses, and Archaea. This field is known as metagenomics, which is defined as the analysis of sequences taken from environmental samples, which are called metagenomes \cite{Wooley2010}. Sequencing these metagenomes produces fragments of sequences, i.e. sequence reads, of organisms that are present in the environmental samples, which may belong to multiple species, and are considered extremely challenging to analyze from a computational perspective. These sequences are usually filtered to exclude those that belong to taxons that are not of interest. The resulting datasets are described using metadata standards, such as MIxS \cite{Yilmaz2011a}, for supporting data discovery and mining. Next, the overlapping sequence reads are used to obtain longer sequences called {\em contigs} in a process known as assembly. Metagenome assembly is often based on de Bruijn graphs \cite{Compeau2011}. Such graphs are built using sequence reads as labels to edges that connect vertices labeled by the prefixes and suffixes of the respective sequence reads. The assembly problem is then reduced to the problem of finding a Eulerian cycle in a de Bruijn graph, i.e. a cycle that visits every graph edge exactly once. Implementations of assemblers that use de Bruijn graphs include SPAdes \cite{Bankevich2012} and HipMer \cite{Georganas2015}. The later is a parallel implementation for high performance computing platforms using the Partitioned Global Adress Space \cite{Yelick2007,Wael2015} programming model. There are several important biological discoveries based on complete and near  complete genomes assembled from metagenomes. For instance, the possible ancestor of mitochondria and the possible ancestor of the first eukaryotic cells were proposed from reconstructed phylogenies from those genomes \cite{Martijn2018DeepAlphaproteobacteria,Eme2017ArchaeaEukaryotes,Zaremba-Niedzwiedzka2017AsgardComplexity}.  After assembly, the resulting sequences can be analyzed for identifying genes using either sequence alignment, for genes with homologs present in public databases, or through {\em ab initio} gene prediction using, for instance, hidden Markov models. Other types of analyses involving metagenomes include evaluation of species diversity and functional annotation \cite{Wooley2010}. Various tools compose these different analyses into metagenomic workflows. MG-RAST \cite{Meyer2008,Wilke2016}, for instance, is a web portal that provides metagenomic dataset analysis workflows containing activities such as quality control, similarity-based annotation, and functional and taxonomic profiling. SUPER-FOCUS \cite{Silva2016} also produces functional and taxonomic profiles from metagenomic datasets. However, its organism identification is based on alignment-free techniques used by the FOCUS \cite{Silva2014} tool. Metagenomics can support various environmental studies such as the analysis of coral diseases. Garcia et al. \cite{Garcia2013} identified taxonomic groups that were more abundant in {\em Mussismilia braziliensis} corals affected by the white plague disease when compared to healthy corals of the same species. Integrating data from metagenomics with data from other aspects of biodiversity, such as species populations and environmental monitoring, is still a challenging task. More recently, there were efforts to integrate these standards \cite{Tuama2012}. Ongoing efforts for improving data integration in bioinformatics and biodiversity are using semantic web techniques \cite{Walls2014}. These efforts are essential in supporting integrative ecosystem studies, such as \cite{Meirelles2015}, where different attributes of the ecosystem found in the mesophotic reefs of the Vitória-Trindade seamount chain were correlated to infer its properties.  

As the computational challenges in this area, we can point questions related to storage, recovery, and integration of the information; conceptual modeling, ontology and semantic representation of the molecular domain. Furthermore, there are usually multiple  computational activities in bioinformatics analyses including filtering, normalization, and annotation. Efforts to ensure  reproducibility \cite{Cohen-Boulakia2017} of these analyses involve (but are not limited to) task composition tools (scripts \cite{Babuji2018ParslPython}, pipelines, scientific workflows \cite{Liew2016a} and software containers \cite{Boettiger2015}), web-based software platforms, such as Galaxy \cite{Bedoya-Reina2013}, commonly used applications, and source code available in repositories such as Github. We explore these issues in more detail in subsection \ref{subsec:swf}.

% ---------------------------------
% --- MACHINE LEARNING WILDLIFE ---
% ---------------------------------

\subsection{Wildlife Health Monitoring}

A comprehensive approach for wildlife health monitoring involves many challenges that need to be addressed in order to result in a globally effective mechanism for diseases prevention, such as:
the difficulty and limited access to wild, mostly uninhabited areas; how to overcome the high diversity and complexity of parasites, vectors, hosts and disease ecology; the methodology and infrastructure for properly collecting, storing and managing georeferenced high-quality data;
how to integrate specialists from different areas to handle data, species and distinct socioenvironmental contexts; the research on knowledge extraction from data-driven models to understand, identify and predict risks to ultimately convey relevant information to society;
and finally, how to sensitize decision-makers about the importance of monitoring as well as to engage the population as committed citizen scientists. The challenges are yet more acute in megadiverse countries which, in addition to biodiversity richness, usually also have to cope with vast territorial distances and sociocultural diversity.

He et al. \cite{He2016} present the eMammal framework for wildlife monitoring supported by citizen scientists. Animal images collected with camera traps are sent to its database where visual animal recognition techniques are applied. The species identification recommendations generated are reviewed by citizen scientists and, subsequently, by experts. The resulting validated records are made available to wildlife and ecological researchers. eBird \cite{Sullivan2014} also leverages the capability of citizen scientists to gather bird observation records. Automated data quality filters are used to support species identifications performed by citizen scientists. In Brazil, there is still the difficulty of building a mechanism that is not impaired by its large territorial extension and its poorly integrated sectoral policies. The Brazilian Wildlife Health Information System (SISS-Geo) \cite{Chame2018} is a platform for collaborative monitoring that intends to overcome the challenges in wildlife health. It aims integration and participation of various segments of society, encompassing: the registration of occurrences by citizen scientists; the reliable diagnosis of pathogens from the laboratory and expert networks; and computational and mathematical challenges in analytical and predictive systems, knowledge extraction, data integration and visualization, and geographic information systems. It has been successfully applied to support decision-making on recent wildlife health events, such as a recent Yellow Fever epizooty \cite{Couto-Lima2017,Moreira-Soto2018}. 

By automating the search for occurrence patterns, the information reaches more efficiently citizens nationwide, from the general population through experts, as well as provides the opportunity for the acquisition of knowledge about the possible patterns and parameters that contribute to the occurrence of diseases. In the medium- and long-term it also builds the capacity of researchers to develop complex modeling in the ecology of diseases that can possibly exploit geographic information in order to improve accuracy. Moreover, occurrence patterns yield data that can assist national policy on health and on biodiversity conservation.

Machine learning has been used for image analysis in wildlife monitoring, such as for automated species classification. As mentioned earlier, in ~\cite{He2016} the authors describe how species recognition is tackled within the eMammal cyber-infrastructure from camera-trap digital images. Once an animal (or group of) crosses the motion sensor and triggers the camera, the resulting sequence of captured images is processed in order to detect and segment the animal from the natural scene. \textit{Detection} is the task of identifying the bounding box within the animals lie on the image, whereas \textit{segmentation} is separating the animals (foreground) from the scene (background). Both tasks are challenging in this context---sometimes even for humans---as the animals are quite camouflaged by the heavy amount of natural elements in the wild. The approach developed to tackle this object-cutting problem~\cite{Ren2013} takes into consideration multiple image frames from the captured sequence: a standard background-foreground classifier is applied to each frame, however, the locally obtained information is fused across all frames in a collaborative manner; this process is repeated iteratively until the refinement converges. According to the authors, this novel technique led to an improvement in the average segmentation precision of near 15\% over the state-of-the-art algorithm. After the background-foreground image segmentation, an even more challenging task takes place: the animal species recognition. Here, the segmented image patches of animals are fed into previously trained machine-learning models---built based on existing labeled images---in order to classify them by species. Since these patches usually contain some elements from the background scene, i.e. the segmentation is not perfect, the recognition model has to be able to cope with a good deal of noise; to make things harder, the model also needs to recognize animals at different poses. Which machine-learning algorithm is the most adequate to train such recognition models depends mainly on the amount of available training data~\cite{Chen2014c}. In summary, conventional supervised classification algorithms are recommended for small training datasets whereas deep-learning algorithms best fit the case of an abundance of labeled data. In~\cite{Chen2014c},
using a training dataset of 14346 images of 20 animal species, a deep convolutional neural network (DCNN) achieved an accuracy of 38\%, while a Bag-of-Words model (BoW) achieved 33\%.
Regardless of the learning algorithm, these species recognition results are still disappointing and of limited use. From an optimistic point of view, though, it is expected that DCNN will perform better with larger training datasets as this architecture is known for its high learning capacity~\cite{Chen2014c}. For instance, in~\cite{Gomez2016} the authors report an accuracy of 88.9\% on a large dataset of near 1 million images containing 26 wild animal species.
Moreover, there is evidence that increasing the deepness of the neural network architecture
leads to higher performance even with no additional training data~\cite{Gomez2016}.

Automatically identifying animals species from images seems to be the current trend in wildlife monitoring. Another great work in this vein was recently presented in~\cite{Norouzzadeh2018},
where the authors, motivated by the need to eliminate the burden of manual labeling by specialists and volunteers, proposed a deep neural network (DNN) to not only identify animals
but also to count them and describe their behavior. As discussed earlier, large datasets are required in order to harness the full potential of DNNs. By using the Snapshot Serengeti (SS) dataset which contains a total of 10.8 million classified images of 48 species to train the deep-learning model, an impressive accuracy of 93.8\% was obtained. The task of recognition was divided into two stages: in the first stage, a model was trained exclusively to separate images containing at least one animal from images without animals. Then, the second DNN model was trained to take the resulting images with animals (only a quarter of the total) and perform the extraction of information, that is, identification of species, number of animals and their characteristics.
Instead of resorting to different models for each task of the second stage, the authors opted for training a single model. The reasoning behind this choice is that (i) learning related tasks simultaneously is more efficient and (ii) a single model has advantages of a reduced amount of parameters and therefore total complexity. In the article, nine DNN architectures were tested,
as well as the ensemble model consisting of all them. For the task of detecting images that contain animals, the best accuracy of 96.8\% was achieved by the Very Deep Convolutional Network architecture (VGG)~\cite{Simonyan2014}. Regarding the species identification, the ensemble of DNNs obtained the highest score, 94.9\%, followed by the Deep Residual Learning architecture (ResNet)~\cite{He2015a}, with 93.8\%. Interestingly, when considered the relaxed definition of accuracy as being \textit{the correct answer in the top-5 guesses by the DNN model}, which would also greatly save human labor in manual labeling, these scores further improve to 99.1\% and 98.8\%, respectively. Counting the number of animals in the image showed to be a hard task for DNNs. Again, the ensemble of DNNs model got the best score, correctly predicting the number of animals 63.1\% of the time. If allowed to approximately count the number of animals up to an error of $\pm 1$, the accuracy climbs to 84.7\%. In this task, the best individual architecture was again ResNet, achieving a score of 62.8\% and 83.6\% for the exact and approximate accuracies, respectively. Finally, the task of describing the characteristics of the animals aimed to detect the following attributes: standing, resting, moving, eating, and whether young animals are present. Note that this is a non-exclusive multilabel classification task, since one or more animals may exhibit multiple attributes. In this task, the ensemble of DNNs model obtained 76.2\% accuracy, and the second best, ResNet, got 75.6\%.

All these enthusiastic results put Deep Learning Networks as the current state-of-the-art when it comes to pattern recognition in wildlife monitoring. This translates to a significant reduction of human labor and also opens a lot of new applications in wildlife monitoring, especially for those small projects that cannot recourse to volunteers and experts for labeling. We are approaching the stage where machine-learning models are as good as humans in species identification, and they can even excel humans in that task, eventually.

\subsection{Other Biodiversity Data Analysis Applications}

The difficulty in the extraction of knowledge from large databases \cite{Han2011}, such as GBIF, demands other methods to access and manage biological data \cite{Howe2008}. As biodiversity databases have observed a substantial increase in data, knowledge extraction from them has become a challenge \cite{Drew2011}. In this sense, Hochachka \cite{Hochachka2007} argues that for the development of ecological analyses where there is little prior knowledge and hypotheses are not clearly developed, exploratory analyses with data mining techniques \cite{Liao2012} are more appropriate than the confirmatory analyses, that is designed to test hypotheses or estimate model parameters. 
In ecology, some research using data mining was conducted by Spehn and Korner \cite{Spehn2009}. Pino-Mejiás et al. \cite{Pino-Mejias2010} used classification algorithms for predicting the potential habitat of species; a decision tree algorithm was also used for forest growing stock modeling \cite{Debeljak2014}; Kumar et al. \cite{Kumar2011} used cluster analysis to identify regions with similar ecological conditions, Flügge et al. \cite{Flugge2014} used multivariate spatial associations for grouping species into disjunct sets with similar co-association values. One of the many possibilities of using data mining was investigated by Silva \cite{Silva2016}, which developed a methodology to allow for the application of association analysis for extracting patterns of co-occurrence from a dataset from the 50ha Forest Dynamics Project on Barro Colorado Island, finding patterns of positive and negative correlation. To do this, association analysis was applied with the Apriori algorithm \cite{Agarwal1994}.  Ciarlegio et al. \cite{Ciarleglio2009} proposed ConsNet, a software for designing conservation area networks using tabu search \cite{Glover1986} with multi-criteria objectives.

% -----------------------------
% --- BIODIVERSITY WORKFLOW ---
% -----------------------------
\subsection{Biodiversity Workflow Management and Reproducibility}
\label{subsec:swf}
Scientific data is being produced at an exponential growth rate by increasingly available scientific sensors. This, coupled with sophisticated computational models that process this data, has demanded new techniques \cite{Hey2009} for managing computational scientific experiments in a scalable and reproducible way. Wilson et al. \cite{Wilson2014} propose best practices for managing scientific computations. These include: recording datasets, programs, libraries, and parameters used, including their respective identifiers or versions, to enable better reproducibility; and using high-level languages for programming and moving to lower-level languages only when performance improvement is necessary. These experiments are often specified as {\em scientific workflows} \cite{Deelman2009,Shade2015,Liew2016a}, which are given by a composition of computational tasks that exchange data through production and consumption relationships. A {\em scientific workflow management system} (SWMS) provides features such as fault-tolerance, scalable execution, scalable data management, data dependency tracking, and provenance recording, that greatly reduce the complexity of managing the life-cycle of these experiments \cite{Mattoso2010}. Scientific workflows are often provided through research data portals, Chard et al. \cite{Chard2018} present a design pattern for such portals for data-intensive scientific problems. Reproducibility \cite{Peng2011} is an essential property in science. In computational research, it can be a challenging task since one might need vasts amounts of data or supercomputing resources to reproduce a result. Sandve et al. \cite{Sandve2013} propose rules that can be followed to better support reproducibility, including recording the steps that were executed to obtain a result, archiving programs that were used in a computational experiment, and versioning the scripts and workflows used. Meng et al. \cite{Meng2015a} propose a framework that tackles reproducibility by providing features for sandboxing and preserving computational environments. A combination of containers \cite{Boettiger2015,Hale2017} and tools for intercepting system calls is used in order to achieve preservation. Many computational experiments have a detailed record of their execution, such as the datasets used and computational tasks used. enable easier verification of results. These records describe the {\em provenance} \cite{Freire2008,Carata2014} of the computational experiment. It can support reproducibility and validation of e-Science experiments. Miles et al. \cite{Miles2007}, for instance, propose an architecture for validation if e-Science experiments based on both provenance assertions and ontologies. DataONE, for instance, included support for tracking the provenance of their datasets \cite{Cao2016DataONE:Support}.

Biodiversity follows the same trend of rapidly increasing production of data found in other areas of science. Currently, biodiversity data is being integrated at a global scale through initiatives such as GBIF \cite{Edwards2000}. Techniques for analysis and synthesis of biodiversity data, such as ENM \cite{TownsendPeterson2011,Elith2009}, are widely used. These analyses typically employ several different applications executed in a loosely coupled manner, being a typical use case for scientific workflow management tools \cite{Liu2015}. 
Next, we list some works related to scientific workflows and reproducibility in biodiversity. Pennington et al. \cite{pennington_ecological_2007} describe the implementation of species distribution modeling (SDM) scientific workflows using Kepler \cite{Ludascher2006}. Their approach allows for easy management of structural aspects of the scientific workflow, such as easily replacing application components. They also developed application components for data transformation and pre-processing, geospatial processing, and semantic annotation of processes.  These experiments use occurrence data from the Mammal Networked Information System (MaNIS)\footnote{http://manisnet.org} and future climatological scenarios from IPCC to predict the climate-change impact on more than 2000 species. Morisette et al. \cite{Morisette2013} present the {\em Software for Assisted Habitat Modeling} (SAHM) that allows for managing the various steps of SDM, including pre- and post-processing activities. The implementation is coupled to the Vistrails \cite{Freire2006} scientific workflow management system, which supports provenance management. Talbert et al. \cite{Talbert2013} also describe SAHM and analyze the data management challenges of SDM using scientific workflows. Amaral et. al \cite{Amaral2014} present the {\em EUBrazilOpenBio Hybrid Data Infrastructure} which implements cloud services for the biodiversity domain such as taxonomic mapping and resolution and SDM. Scientific workflows are supported both with DAGMan \cite{Couvares2007} and EasyGrid AMS \cite{Boeres2004}. They evaluate the execution of SDM on cloud computing resources showing good performance. Candela et al. \cite{Candela2013} give a detailed description of an integrated cloud-based environment for SDM of the of EUBrazilOpenBio Hybrid Data Infrastructure which includes components for retrieving species occurrences, environmental layers, and execution of various models for predicting species distributions. Some SDM applications and workflows are available through web portals, such as  the Biodiversity Virtual e-Laboratory (BioVel) \cite{Hardisty2016}. BioVel \cite{Hardisty2016} offers a web-based environment for managing scientific workflows  for biodiversity. Various pre-defined activities are available in its interface: geographic and temporal selection of occurrences (BioSTIF), data cleaning, taxonomic name resolution, ecological niche modeling algorithms,
 population modeling, ecosystem modeling, and metagenomics and phylogenetics applications \cite{Vicario2012}.

iPlant \cite{Goff2011} is a computational research infrastructure, or cyberinfrastructure, for plant science. Its applications include the {\em Tree of Life} to produce phylogenetic trees of all green plant species; and {\em Genotype to Phenotype} to predict plant phenotypes from their genetic data. Kurator \cite{Dou2012} is a software package for the Kepler \cite{Ludascher2006} scientific workflow management system that supports composing various data curation activities into scientific workflows. Pre-built activities include georeferencing, scientific name, and flowering time validators. Provenance is recorded to document all the transformations activities that data went through caused by the various data curation activities. Nguyen et al. \cite{Nguyen2017} developed scientific workflows for assessing ecosystem risk based on IUCN guidelines \cite{Keith2013} that use five rule-based criteria to assign one of eight risk categories that range from {\em least concern} (LC) to {\em collapsed} (CO). The assessment is performed in two phases. First, a stochastic ecosystem model is executed for the Meso-American Reef Ecosystem risk assessment by predicting future reef properties under diverse scenarios. This step was implemented both in Nimrod/G \cite{Abramson2000} and Spark \cite{Zaharia2016}, for comparative purposes. The Spark version had a considerably better performance in terms of computing time. Next, a workflow was implemented in the Kepler scientific workflow management system \cite{Ludascher2006} to execute the IUCN ecosystem risk assessment methodology using the results of the stochastic ecosystem model execution and applying its five rule-based criteria. Reproducibility is an important property in this process since risk assessment is often re-executed and its results need to be discussed by experts and decision makers \cite{Guru2016}.

Cohen-Boulakia et al. \cite{Cohen-Boulakia2017} explore the use of scientific workflows for reproducibility of computational experiments in the life sciences. They analyze scientific workflows techniques and systems and evaluate to what extent they support reproducibility requirements in life science applications. Plant phenotyping, which evaluates how plants respond to different environmental conditions by monitoring their traits, was one of the use cases. For instance, keeping track of several different tool versions used in a workflow and their respective compatibility is one of its reproducibility requirements. They define different levels of reproducibility in the workflow context. Considering two scientific workflows $A$ and $B$ and assuming $A$ has already been executed. When $B$ is executed: {\em repeatability} is achieved when $B$ contains exactly the same components of $A$; {\em replicability} is obtained when $B$ uses similar \cite{Starlinger2014} input components of $A$ and both executions reach the same conclusion; {\em reproducibility} happens when both executions lead to the same scientific conclusion;  {\em reusability} is observed when the specification $B$ contains the specification of $A$. The authors analyze three workflow aspects from the reproducibility perspective: workflow specification, workflow execution, and workflow context and runtime environment. Workflow specifications can support better reusability through common specification languages, such as CWL\footnote{\url{http://www.commonwl.org/}} (Common Workflow Language), and annotations. Assessing workflows similarity is critical for reuse but progress is still needed in solving this problem. Recording and analyzing workflow execution details can be supported by provenance information. While most systems support the PROV \cite{Moreau2015} standard, visualizing and analyzing large provenance datasets is still challenging. Also, preserving the runtime environment is still a challenge that is being addressed with virtualization technologies \cite{Hale2017}. WholeTale \cite{Brinckman2018}, for instance, is a computational environment that has reproducibility features. It has components for data collection, identity management, data publication, and interfaces to analytical tools, called frontends. These frontends will manipulate data and can be given, for instance, by interactive notebooks such as Jupyter\footnote{\url{http://jupyter.org}}. The system is integrated with DataONE \cite{Michener2012}, users can search and retrieve datasets from it. Frontends are packaged as Docker containers \cite{Boettiger2015} that can be executed on high performance computing resources. The interaction between the datasets and analytical tools is documented and recorded in a metadata management system. This allows for reproducing the entire computational research performed, from data retrieval to data analysis and its outputs, including the computational environments used.

% ******************************
% ***** CONCLUDING REMARKS *****
% ******************************
\section{e-Biodiversity Challenges and Concluding Remarks}
\label{sec:conclusion}

The acceleration of global changes requires a constant assessment of their impacts on biodiversity and, consequently, on ecosystem services that are essential to human survival. Some areas of the globe, e.g. the south Atlantic Ocean, remain highly understudied and therefore their biodiversity underestimated. A better understanding of marine biodiversity could be achieved with help of e-Biodiversity to leverage surveys to uncover novel species and systems, such as the Great Amazon Reef \cite{Francini-Filho2018PerspectivesThreats}. To address this problem, biodiversity data must be systematically collected and analyzed. In this context, e-Biodiversity, or Biodiversity Informatics, is an essential collection of methodologies, tools, and techniques to achieve this goal. The EBVs \cite{Pereira2013} were proposed as a set of indicators that would allow for systematic monitoring of biodiversity. However, the production of these indicators is still a challenge \cite{Peterson2017}, in particular regarding the existence of information gaps that can prevent global-scale inferences on the state of biodiversity. These inferences provide essential input to decision-makers in devising governmental policies toward meeting global targets on biodiversity conservation, such as the Aichi Biodiversity Targets. In this section, we describe existing challenges for e-Biodiversity to become a systematic and global-scale tool for monitoring and making inferences about biodiversity. 

As described in subsection \ref{sec:dqffu}, the availability of detailed information about most organisms is still very scarce \cite{Peterson2006UsesModels}. 
This hinders the usage of this data in many biodiversity data analysis applications, such as ecological niche modeling. Furthermore, we described other issues with biodiversity data, such as biases and frequent taxonomic and georeferencing errors. Therefore, one of the challenges of e-Biodiversity is not only to increase the amount of available data, filling some of the existing gaps but also to reduce its bias and improve its quality. Some promising work addressing these issues are listed next:
\begin{itemize}
\item Heidorn \cite{Heidorn2008} observes that data from smaller scientific projects is rarely available to other scientists even though its aggregated size and value for research is considerable. This phenomenon is denominated the {\em long tail of science}. In biodiversity and ecology, some progress has been achieved through projects such as GBIF and DataONE, that receive a considerable amount of their datasets from small research groups. One issue with making these datasets available is the effort required to map the concepts present in them to standard vocabulary terms used in major biodiversity databases. Entity resolution techniques \cite{Kopcke2010EvaluationProblems} have the potential to assist and speed up these record linkage routines.
\item Data collection could be substantially intensified by applying artificial intelligence methods for automating specimen identification, some preliminary work in this direction include the application of deep learning techniques for species identification in herbarium sheets \cite{Carranza-Rojas2017,Carranza-Rojas2018}. 
\item Remote sensing provides the opportunity to regularly observe the Earth and, therefore, could benefit biodiversity monitoring by increasing the amount of data collected. It can also be a valuable tool to observe areas that are difficult to access through field expeditions. One of the pioneering works in this area was proposed by Holden and Ledrew \cite{Holden1999HyperspectralFeatures} by using hyperspectral remote sensing to monitor coral reefs. Clark et al. \cite{Clark2005} identified tree species  using remote sensing images. Fretwell et al. \cite{Fretwell2012AnSpace} were able to use satellite-based remote sensing to survey the Emperor Penguin on a global scale.  
\item As observed in section \ref{sec:dqffu}, assessing the quality of a dataset is a critical step for any subsequent analysis and synthesis activity that might use it. Users should establish the intended use of datasets in their research. Determining if a dataset is fit for a particular use is still a challenge in e-Biodiversity since records available in public databases contain various types of errors. A promising approach was proposed by Veiga et al. \cite{Veiga2017} comprised of a framework for biodiversity data quality assessment and management that allows for users to define their data quality requirements and when a particular dataset is fit-for-use in a standardized manner. Morris et al. \cite{Morris2018} made some progress by implementing a library of small data quality assessment routines that can be composed into more complex workflows, to report data quality in terms of the framework proposed by Veiga et al. \cite{Veiga2017}.
\end{itemize}

In the Global Biodiversity Information Outlook \cite{hobern_global_2013} report, produced by leading biodiversity informatics researchers, a number of areas of biodiversity informatics of limited or minimal progress were identified and can be considered research challenges. Biological system modeling was considered an area of research in biodiversity informatics with minimal progress. Advances in this area could be comprised of computational, or {\em in-silico} models or simulations ranging from single organisms to entire ecosystems. Current temporal and spatial modeling in biodiversity, such as ecological niche modeling, described in subsection \ref{subsec:enm}, rely on species occurrence data. More fine-grained modeling would require incorporating species trait data. Cardinale et al. \cite{Cardinale2012}, for instance, advocated the development of new predictive models that take into account species interactions to predict the impact of biodiversity on ecosystem processes based on species traits. Areas of limited progress identified by the Global Biodiversity Information Outlook included:
\begin{itemize}
\item Automated remote-sensed observation has the potential to enable observation of biodiversity in large and remote areas. More recently, preliminary work was conducted on defining biodiversity indicators that could be derived from images collected by satellite remote sensing \cite{Pettorelli2016} and processed using statistical analysis and classification algorithms. 
\item Identifying trends and making predictions about biodiversity could determine future trends in biodiversity under different global change scenarios. Ongoing research in this area includes predicting how climate change will affect species distributions  \cite{Siqueira2003,Thomas2004,Pearson2006,Araujo2008,Wiens2009,Araujo2012} and zoonotic diseases \cite{Estrada-Pena2014}.
\item Providing access to aggregate species trait data, consisting of data on species characteristics and their interactions. Data is incomplete, there is not much data about relative abundances of species, their traits, and on how they interact. This information is needed for creating better models to study ecosystem processes. This type of data can enable more complex biological systems modeling, such as evolutionary inference. MorphoBank \cite{OLeary2011MorphoBank:cloud}, for instance, allows for scientists to upload images with the morphology of organisms with associated data.
\end{itemize}

Scientists often need to combine multiple sources of data \cite{Jones2006,Fujioka2014} in biodiversity analysis and synthesis activities. Although there are many gaps in biodiversity data, such as the reduced availability of species trait data, there are many machine-readable and freely available, i.e. open, datasets \cite{Reichman2011} from areas such as remote sensing, socioeconomics, and climatology, that can be integrated into biodiversity studies. Open data is widely available online, including data provided by many governments. However, it is highly heterogeneous, dispersed in multiple sources, and may not provide metadata or schema. Metadata, described in section \ref{sec:metadata}, is helpful in discovering datasets and in integrating them when dataset attribute definitions are provided, as it is possible with EML \cite{Fegraus2005a}. Semantic web \cite{Walls2014}, as described in section \ref{sec:ddi}, can also support data integration through the use of various existing ontologies for biodiversity and other domains. However, their increased usefulness depends on the widespread adoption of ontologies and metadata standards by data providers, a process that is still underway. A promising approach to overcome these limitations has been to use machine learning techniques to support open data integration activities \cite{Miller2018OpenIntegration,Dong2018DataLearning}, such as entity matching \cite{Mudgal2018DeepMatching,Nargesian2018TableData}. These recently proposed techniques could be leveraged and extended for integrating biodiversity and other related datasets.

Trends, indicators, and facts derived from biodiversity data analysis and synthesis activities might be used for guiding governmental decision making in critical areas such as conservation area planning, impact assessment of large construction work projects, and zoonotic disease prevention. Since such decisions can have large-scale impacts in society, being able to trace back the processes involved in reaching them is an essential property. Therefore, it is important to use methodologies and techniques that are reproducible \cite{Peng2011,Sandve2013,Cohen-Boulakia2017,Ivie2018ReproducibilityComputing} when executing these activities. Some initial advances were achieved in projects such as DataONE \cite{Cao2016DataONE:Support} and WholeTale \cite{Brinckman2018} by recording the provenance of biodiversity datasets and of their analysis and synthesis. However, reproducibility in computational science, in general, is still a challenge. Reproducibility tool usability is still considered low \cite{Freire2018} making it difficult or time-consuming for scientists to keep track of the data derivation steps in their analyses. Another challenge in reproducing a scientific workflow is finding similar scientific workflows and verifying if they reach similar conclusions \cite{Cohen-Boulakia2017}. Starlinger et al. \cite{Starlinger2016EffectiveRepositories} advanced in addressing this problem by proposing a layer decomposition of scientific workflows into an ordered representation of its activities suitable for comparison. Providing reproducible frameworks for biodiversity analysis and synthesis activities would enable better decision traceability and validation of trends and indicators produced by them.

\section*{Acknowledgements}

The work is partially supported by CAPES, CNPq, and FAPERJ.

% ==========================
% %%%%%% Bibliography %%%%%%
% ==========================
\footnotesize
\bibliographystyle{plain}

\end{document}